\documentclass[a4paper,fleqn,usenatbib]{mnras}

\usepackage{aasmacros}
\usepackage{times}
\usepackage{amsmath}
\usepackage{graphicx}
\usepackage{amssymb}
\usepackage[usenames,dvipsnames]{color}
\usepackage{url}
\usepackage{fixltx2e}
\usepackage{mathtools}
\usepackage{amsmath}
\usepackage{float}
\usepackage{cite}
\usepackage{fixltx2e}
\usepackage{longtable}
\usepackage{hyperref}
\usepackage{morefloats}
\usepackage{totpages}
\usepackage{amssymb}
\usepackage{xcolor}
\usepackage{enumitem}
\usepackage{tabu}
\usepackage[T1]{fontenc}
\usepackage{ae,aecompl}
\usepackage{mathptmx}
\usepackage{txfonts}

\let\cite=\citen

\newcommand\T{\rule{0pt}{2.6ex}}       % Top strut
\newcommand\B{\rule[-1.2ex]{0pt}{0pt}} 

\newcommand{\trm}{\textrm}

\def\apj{ApJ}
\def\aj{AJ}
\def\apjs{ApJl}
\def\apjl{ApJl}
\def\mnras{MNRAS}

\def\pasp{PASP}

\def\aspcs{ASPCS}

\def\speak{$S_{\rm peak}$}
\def\thetamaj{$\theta_{\rm maj}$}
\def\thetamin{$\theta_{\rm min}$}
\def\posang{$\phi$}
\def\thetabeam{$\theta_{\rm B}$}
\def\stot{$S_{\rm total}$}
\def\thetadmaj{$\theta_{\rm Dmaj}$}
\def\thetadmin{$\theta_{\rm Dmin}$}

\def\Ob{\textsc{Obit}}

\makeatletter
\newcommand*{\rom}[1]{\expandafter\@slowromancap\romannumeral #1@}
\makeatother

\def\OW{Owen \& Morrison } 
\defcitealias{Condon12}{CO12}
\defcitealias{Vernstrom16a}{Paper \rom{1}}
\bibpunct{(}{)}{;}{a}{}{,}

\title[Deep 3-GHz VLA catalogue]{Deep 3-GHz Observations of the Lockman Hole North with the Very Large Array -- \rom{2}. Catalogue and $\boldsymbol{\mu}$Jy source properties}

\author[Vernstrom et. al]{T. Vernstrom\thanks{E-mail:vernstrom@dunlap.utoronto.ca}$^1$, Douglas Scott$^2$, J.V. Wall$^2$, J.J. Condon$^3$, W.D. Cotton$^3$,\newauthor
 K.I. Kellermann$^3$, R.A. Perley $^4$ \\
  $^1$Dunlap Institute for Astronomy and Astrophysics, University of Toronto, Toronto, ON M5S 3H4, Canada\\
  $^2$Department of Physics and Astronomy, University of British Columbia, Vancouver, BC V6T 1Z1, Canada\\
  $^3$National Radio Astronomy Observatory, 520 Edgemont Rd, Charlottesville, VA 22903, USA \\
  $^4$National Radio Astronomy Observatory, P.O. Box 0, Soccoro, NM 87801, USA\\
}
\begin{document}

\pagerange{\pageref{firstpage}--\pageref{lastpage}} \pubyear{2015}

\maketitle

\label{firstpage}
\begin{abstract}
This is the second of two papers describing the observations and source catalogues derived from sensitive $3$-GHz images of the Lockman Hole North using the Karl G. Jansky Very Large Array. We describe the reduction and cataloguing process, which yielded an image with 8 arcsec resolution and instrumental noise of $\sigma_{\rm n}=1.01\,\mu$Jy beam$^{-1}$ rms (before primary beam corrections) and a catalogue of 558 sources detected above $5\sigma_{\rm n}$. We include details of how we estimate source spectral indices across the 2-GHz VLA bandwidth, finding a median index of $-0.76\pm0.04$. Stacking of source spectra reveals a flattening of spectral index with decreasing flux density. We present a source count derived from the catalogue. We show a traditional count estimate compared with a completely independent estimate made via a {\it P(D)} confusion analysis, and find very good agreement. Cross-matches of the catalogue with X-ray, optical, infrared, radio, and redshift catalogues are also presented. The X-ray, optical and infrared data, as well as AGN selection criteria allow us to classify 10 per cent as radio-loud AGN, 28 percent as radio-quiet AGN, and 58 per cent as star-forming galaxies, with only 4 per cent unclassified. 
\end{abstract}


\section{Introduction}
\label{sec:introduction}
The sub-mJy radio source population at frequencies near $1.4\,$GHz contains a mixture of radio-loud and radio-quiet active galactic nuclei (AGN) and star-forming and starburst galaxies \citep[e.g.][]{Condon89,Afonso05,Simpson06,Bonzini13,Padovani15}. Both classes of galaxies emit synchrotron radiation, but the emitting cosmic-ray electrons are accelerated by the accreting black hole of the AGN in the first case and by supernova remnants of stars with $M> 8 {\rm M_{\odot}}$ in the second. These massive stars have lifetimes $\lesssim 3 \times10^8\,$years, so their supernova rates are proportional to the recent star-formation rates of their host galaxies. Radio galaxies and radio-loud AGN dominate at flux densities greater than about $1\,$mJy, and have therefore been historically easier to study. 

In the $\mu$Jy regime (below roughly $1\,$mJy), star-forming and starburst galaxies become dominant \citep{Prandoni01a,Huynh08,Seymour08,Smolcic08,Mao12,White12}, as well as radio-quiet AGN \citep{Padovani09,Padovani11,Padovani15}. The radio-quiet AGN show the presence of AGN activity in one or more bands of the electromagnetic spectrum, but the radio emission itself is powered by star-formation processes \citep{Bonzini15}. Furthermore, whether one classifies a source as AGN or star-forming depends on the criterion adopted. Understanding the characteristics of this mixed population of $\mu$Jy radio sources is important for the study of the star-formation history of the Universe \citep{Karim11}, the cosmic AGN activity history \citep{Smolcic15}, and for the planning of future deep radio surveys, such as with the Square Kilometer Array (SKA).

The issue is complicated by the existing $1.4$-GHz source count data in this $\mu$Jy range, where there is considerable scatter among published surveys \citep[see e.g.][]{Condon07,Dezotti09}, more than expected from ``cosmic'' or sampling variance \citep{Heywood13}. This is likely due to the different survey systematics and corrections applied to the raw observations. \citet{Condon12} and \citet{Vernstrom13} showed that the confusion analysis {\it P(D)} approach can be used to obtain estimates of the source count well into the $\mu$Jy range. To break up the count into different population components, multi-wavelength data are needed \citep[e.g.][]{Padovani11}. Therefore, sensitive radio observations, along with deep multi-wavelength data, combined with a careful investigation of the associated uncertainties, are needed to further our understanding of faint radio source populations. 

\citet{Vernstrom16a} (hereafter \citetalias{Vernstrom16a}) describes the uncertainties involved in the cataloguing process. In this paper we present a new radio source catalogue and source characteristics, from sensitive observations of the extragalactic region of extensive study known as the Lockman Hole North, using the Karl G. Jansky Very Large Array (VLA) at $3\,$GHz. The outline for this paper is as follows.  In Section~\ref{sec:obs} we briefly describe the data obtained from the VLA and the radio image characteristics. Section~\ref{sec:cat_cat} describes the catalogue method and results, including the angular size distributions. In Section~\ref{sec:cat_count} we present the source count. Section~\ref{sec:cat_si} gives a discussion of spectral indices derived from the $2$-GHz bandwidth. Finally, Section~\ref{sec:cat_crxid} describes the procedure and results of cross-matching the radio catalogue with catalogues at other wavelengths. 

\section{Observations}
\label{sec:obs}

The observations were made with the VLA at $S$-band, which spans from approximately $2\,$GHz to $4\,$GHz. There were two main observing sessions: one with the C configuration, with an average of 21 antennas; and one with higher resolution in the BnA and B-to-A configurations. The BnA configuration is a hybrid of the B and A configurations, while B-to-A refers to data taken during the transition from the BnA to A configuration. Even though the BnA observations included baselines longer than the longest B-configuration baseline, we weighted the data so that the resultant resolution is comparable to that of the smaller, or B, configuration. The observing dates and time spent in each configuration are listed in Table~\ref{tab:configs}. There was a total of approximately $57$ hours of on-source observing time in the C configuration, and 24 hours in the BnA and B-to-A configurations.

\begin{table}
\centering
\caption{VLA Observing runs summary. The B$^{*}$ data were taken during BnA configuration and transition period to A configuration; however, images were made at B configuration resolution.} 
\label{tab:configs}
\begin{tabular}{lcccc}
\hline
\hline
Array & Start date & End date & Hours&No. of sessions\\
\hline
C & 2012 Feb 21 & 2012 Mar 18& 57.0 & 6\\
B$^{*}$ &2014 Feb 02 & 2014 Feb 17&  24.0 & 9\\
\hline
\end{tabular}
\end{table}

The $3$-GHz VLA field was selected explicitly to overlap the region Owen $\&$ Morrison (2008) observed in the Lockman Hole at $1.4\,$GHz. The field is centred on $\alpha = 10^{\rm{h}}46^{\rm{m}}00^{\rm{s}}$, $\delta = +59^{\circ}01'00''$ (J2000), and was originally chosen as it lacks many bright radio sources, with the brightest being about $7\,$mJy, and no other very bright radio sources nearby. It is also covered in many other wavebands (with {\it Spitzer, Chandra, Herschel}, GMRT, and more) allowing for source cross-identification, investigation of AGN contribution, and study of the far-IR/radio correlation. The $3$-GHz (centre frequency) S-band was chosen rather than the $1.4$-GHz L-Band, because with the upgraded VLA the contamination from interference is less, the requirements on dynamic range are lower, the confusion is lower, and S-Band has greater available bandwidth ($2 \, $GHz).

The calibration, editing, and imaging were performed using the \textsc{Obit} package \citep{Cotton08},\footnote{\url{http://www.cv.nrao.edu/~bcotton/Obit.html}} and are described in detail in \citet{Condon12}.\footnote{\citet{Condon12} describe the details of the C-configuration data; however, the steps were the same for the BnA configuration data, which were added to the C configuration data once they were calibrated, and the combined data imaged.} The VLA S-band contains 16 frequency sub-bands. An image of each sub-band was made with {\it uv} weighting applied to ensure each sub-band had the same size and shape PSF, with the size predominately dictated by the lowest frequency bands. This resulted in circular $8\,$arcsec synthesized beams for the C-configuration-only data (herein ``C data" or ``C image"), and the C and BnA configurations combined (herein ``CB data" or ``CB image") having circular $2.75\,$arcsec synthesized beams. For the CB images this weighting includes the weighting (discussed above) to achieve a resolution closer to that of the B-configuration. The sub-band images were imaged, or CLEANed, simultaneously. The \textsc{AIPS} task\footnote{\url{http://www.aips.nrao.edu}} \textsc{IMEAN} was used to calculate the rms noise, $\sigma_{\rm n}$, of the CLEANed sub-band images in several large areas well outside the main lobe of the primary beam and containing no visible ($S_{\rm peak} \ge 6\,\mu{\rm Jy~beam}^{-1}$) sources. The $\sigma_{\rm n}$ values are listed in Table~\ref{tab:obs_subb}.\footnote{The BnA-configuration data had a few sub-bands with much higher noise, as is evident in the $\sigma_{\rm n}$ values shown in Table~\ref{tab:obs_subb}. This is because very few of these data survived the editing and calibration process. For imaging and analysis these sub-bands were given very low weights when combining the images. } The FWHM of the primary beam ranges from $21.6\,$arcmin at $2\,$GHz to $ 10.8\,$arcmin at $4\,$GHz. 

\begin{figure}
\includegraphics[scale=.5,natwidth=6in,natheight=12in]{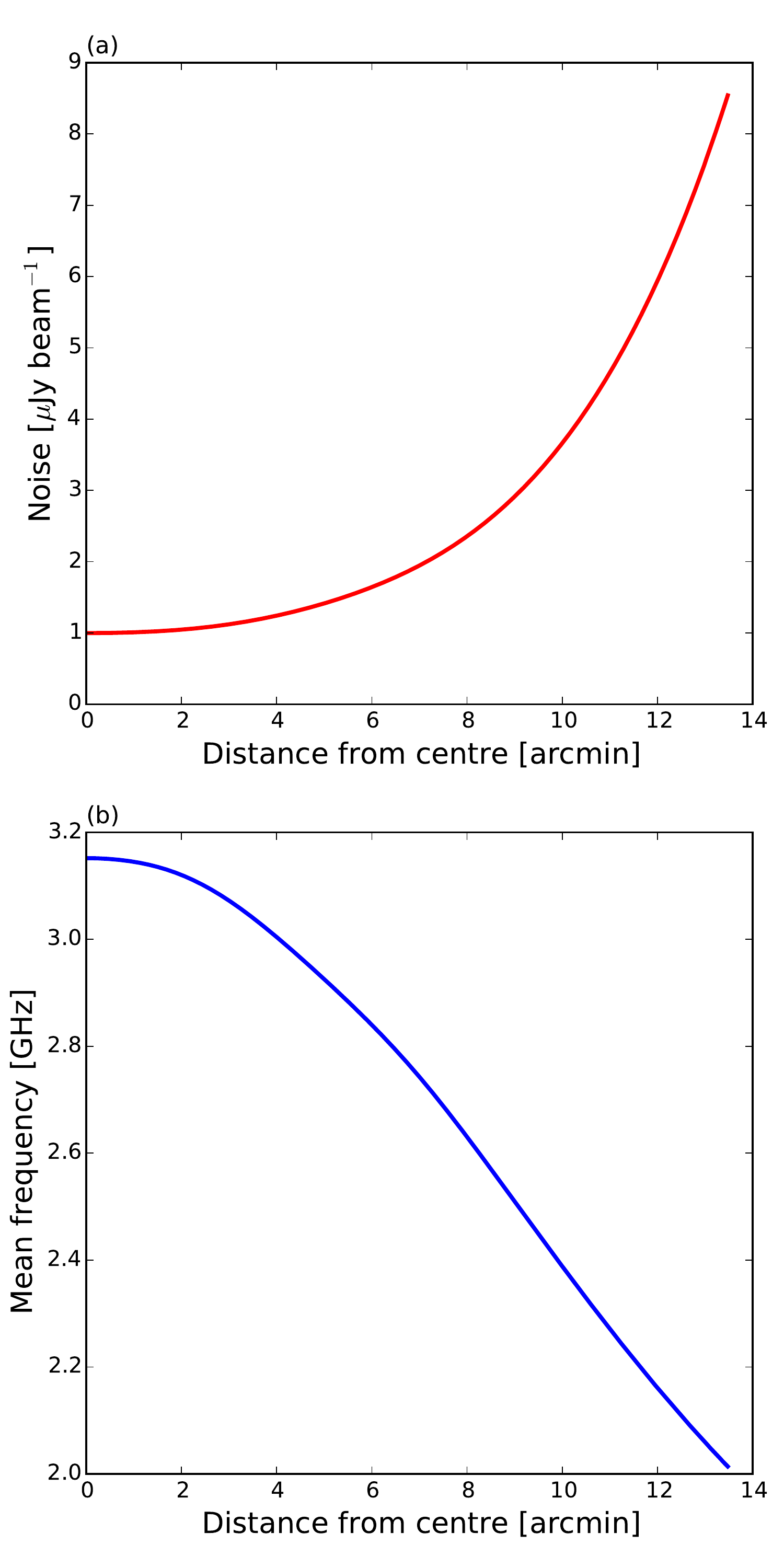}
\caption{Radial profiles of VLA images. Panel (a) shows the change in the noise (of the C data) as a function of distance from the image centre, after correction for the primary beam. Panel (b) shows the mean frequency as a function of distance, after wide-band combination and primary beam correction.  }
\label{fig:sigfreq}
\end{figure} 

\begin{figure}
\centering
\includegraphics[scale=.33,natwidth=10.5in,natheight=9in]{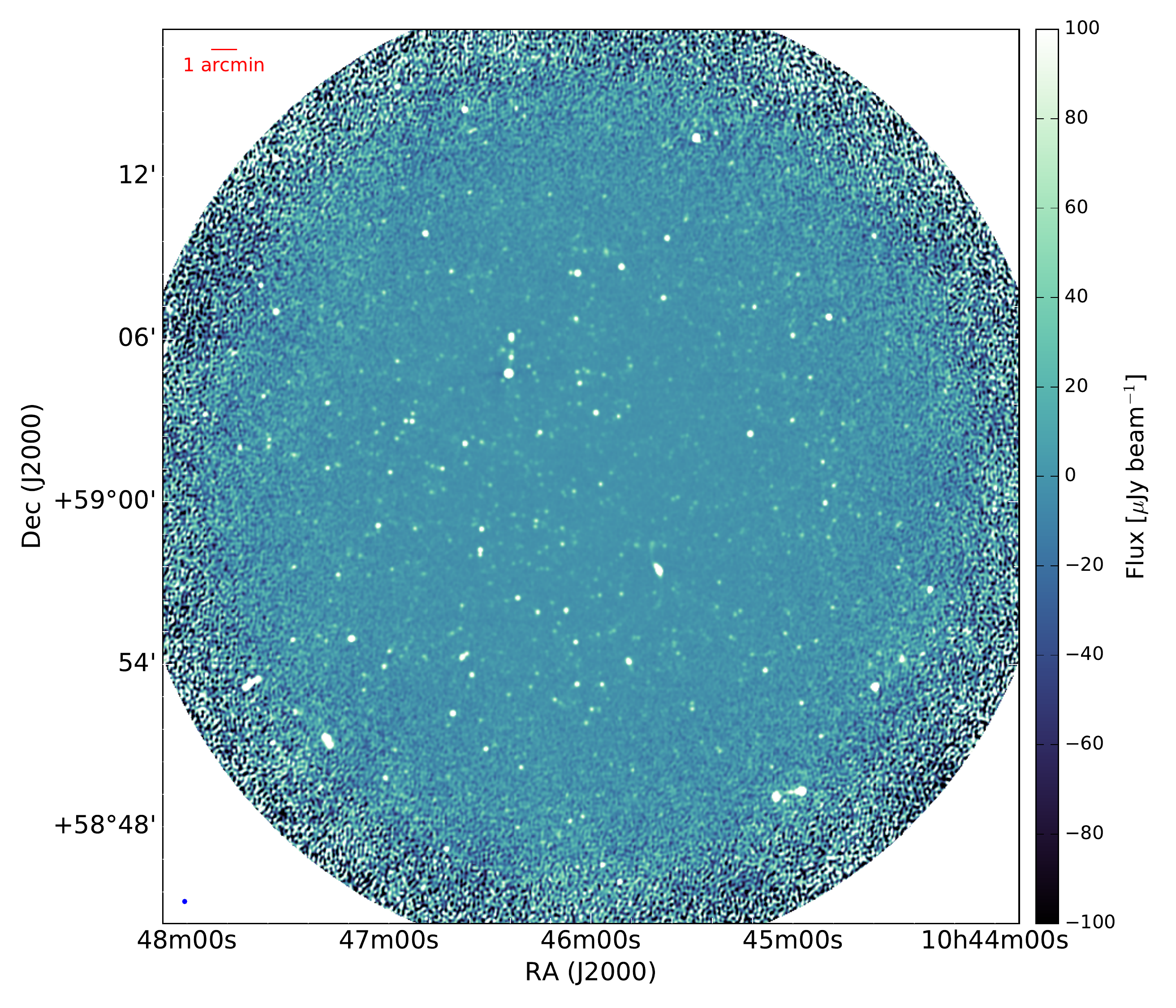}
\caption{VLA 3-GHz Lockman Hole C-image after correction for the primary beam. The size of the field shown here extends to the $10\,$per cent power radius of the primary beam.}
\label{fig:vlaims}
\end{figure}

To combine the sub-bands each was weighted to maximize signal-to-noise for sources having spectral indices near $\alpha = -0.7$, the mean spectral index of faint sources found at frequencies around $3 \,$GHz \citep{Condon84b}. We assigned to each sub-band image pixel a weight inversely proportional to its primary-beam-corrected noise variance $\sigma_{\rm n}^2$ scaled with $\alpha$, and generated a corrected wide-band image from the weighted average of the sub-band images. The primary-beam-corrected C image is shown in Fig.~\ref{fig:vlaims}. 

\begin{figure}
\includegraphics[scale=.38,natwidth=8.34in,natheight=7.64in]{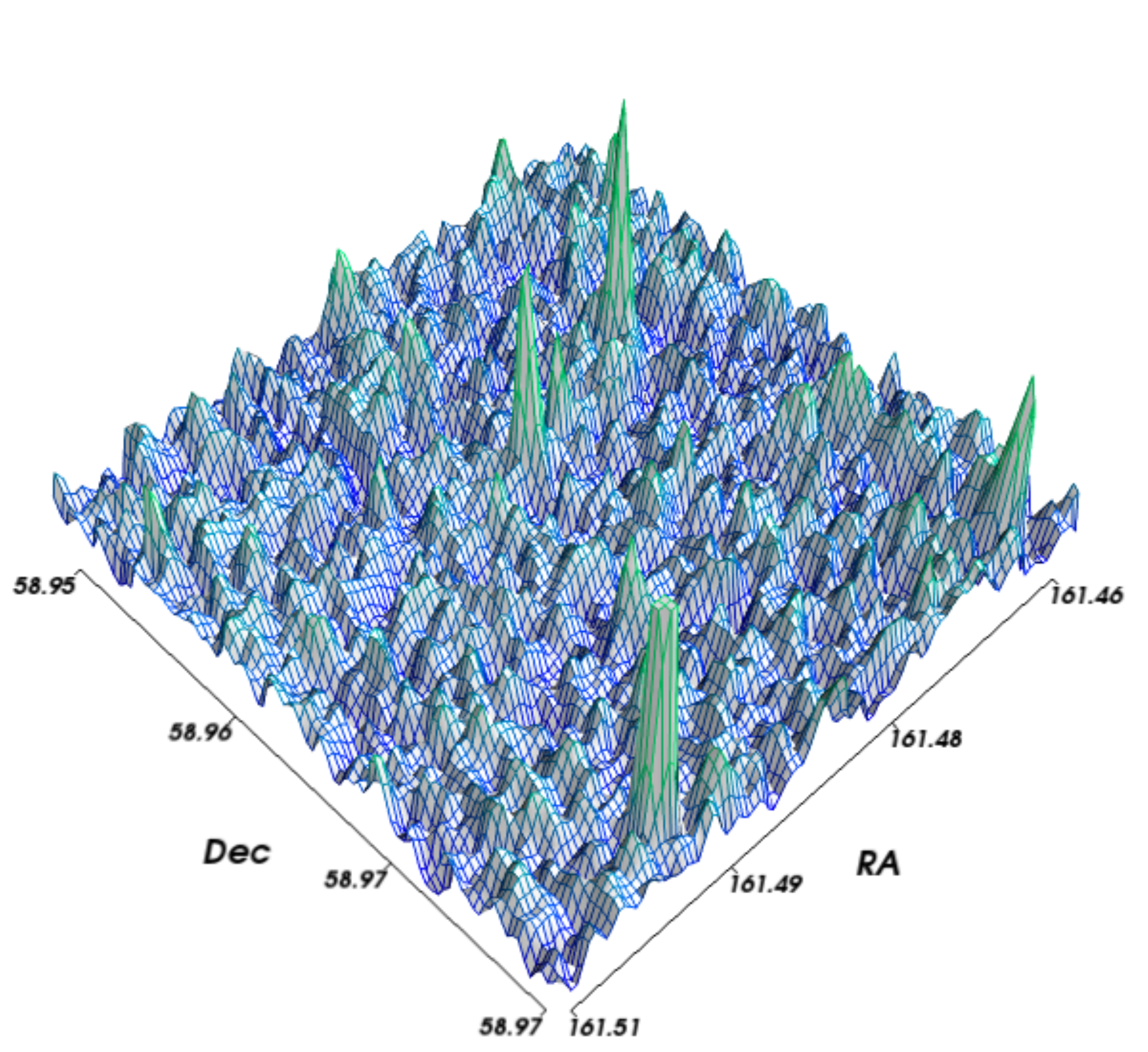}
\includegraphics[scale=.38,natwidth=8.34in,natheight=7.64in]{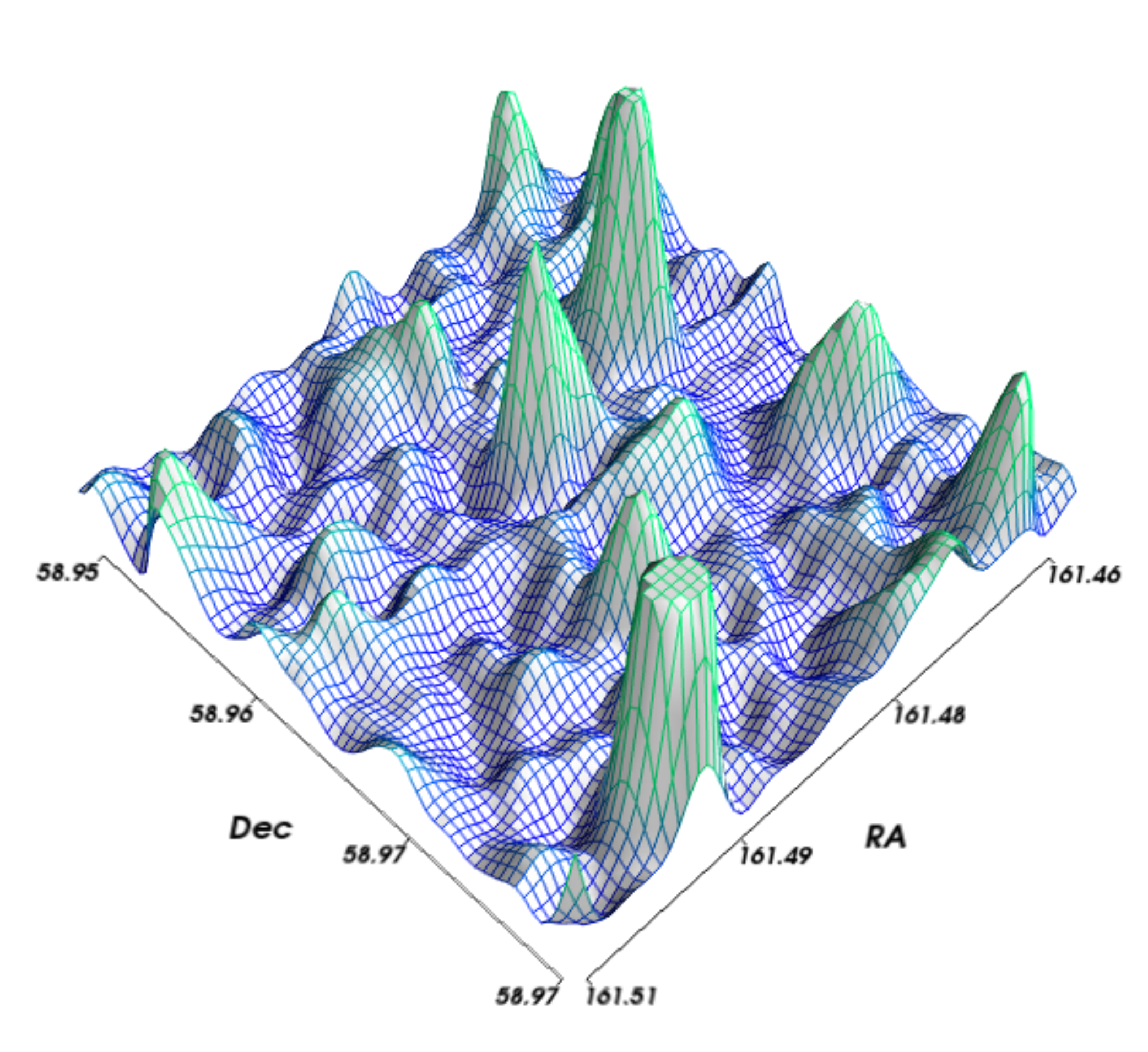}
\caption{Three-dimensional representations of confusion profiles of a small region of the CB image (top) and C image (bottom). This shows that many of the individual peaks in the CB data are blended together in the C-image.  }
\label{fig:im3d}
\end{figure}

We measured the noise distribution in four large regions well outside the primary main beam and free of visible sources. The best-fit to the noise indicates a Gaussian distribution with an rms of $\sigma_{\rm n} = 1.01 \,\mu{\rm Jy\,beam}^{-1}$ for the C data and $\sigma_{\rm n} = 1.15 \,\mu{\rm Jy\,beam}^{-1}$ for the CB data, before primary beam corrections. The radial dependence of the noise after primary-beam correction is shown in panel (a) of Fig.~\ref{fig:sigfreq}. The CB image has higher instrumental noise than the C image due to the lower weight given to the shorter baselines which dominate the C data relative to that given the longer baselines. While the C image may have lower instrumental noise, it has higher confusion noise (i.e. more blending of sources due to the larger beam). This can be seen from a comparison of a small region of each shown in Fig.~\ref{fig:im3d}, where many of the small spikes in the CB image are blended in the C image. 

Because of the weighting used to combine the individual images, the primary beam of the wide-band image is frequency-dependent, so the effective frequency $\langle\nu\rangle$ of the image decreases with radial distance from the pointing centre. The radial dependence of the mean frequency is shown in panel (b) of Fig.~\ref{fig:sigfreq}. Table~\ref{tab:images} provides a summary of the final image properties.

 \begin{table}
 \centering
 \caption{Frequency and noise properties of the 16 VLA sub band images.}
  \label{tab:obs_subb}
 \begin{tabular}{cccc}
\hline
\hline
  Sub band & $\nu_{\rm c}$&C data $\sigma_{\rm n}$ & CB data $\sigma_{\rm n}$ \\ 
number & [GHz]& [$\mu{\rm Jy~beam}^{-1}$]& [$\mu{\rm Jy~beam}^{-1}$] \\
\hline
01 & 2.0500 &  9.22  & 8.46\\ 
02 & 2.1780 &  {\llap 1}3.19& {\llap2}6.1 \\ 
03 & 2.3060 &  {\llap 1}8.17 & {\llap 100}0\\ 
04 & 2.4340 & 4.48 & 5.09\\ 
05 & 2.5620 & 4.24 &5.00\\ 
06 & 2.6900 & 4.31 &4.42\\ 
07 & 2.8180 &  4.04 &4.26\\ 
08 & 2.9460 &  3.72 &4.75\\ 
09 & 3.0500 & 4.48 &4.01\\ 
10 & 3.1780 & 3.14 &3.49\\ 
11 & 3.3060 & 3.07 &3.56\\ 
12 & 3.4340 & 2.97 &3.19\\ 
13 & 3.5620 & 2.88 &3.47\\ 
14 & 3.6900 & 3.47 &4.04\\ 
15 & 3.8180 & 4.45 &5.66\\ 
16 & 3.9460 & 4.40 &{\llap117}\\ 
\hline
\end{tabular}
\end{table}

\begin{table}
\centering
\caption{Image properties for the wide-band VLA data. The reported noise values are all after correction for the primary beam and frequency weighting effects, with $\rho$ being the distance from the pointing centre. The clean beam size, $\theta_{\rm B}$, is the FWHM, and the field of view, (FOV), is the FWHM of the primary beam at $3\,$GHz.   }
\label{tab:images}
\begin{tabular}{llll}
\hline
\hline
Quantity & C-data & CB-data& Unit \\
\hline
$\langle\nu\rangle (\rho=0)$ & 3.06 & 3.036&GHz\\
$\langle\nu\rangle(\rho=5^{\prime})$ & 2.96 &2.96& GHz\\
Pixel size & 0.5 & 0.5&arcsec\\
$\theta_{\rm B}$ & 8.00 &2.75& arcsec\\
FOV & 13.75 &13.75&arcmin \\
$\sigma_{\rm n}(\rho=0)$ & 1.01 &1.15& $\mu$Jy beam$^{-1}$ \\
$\sigma_{\rm{n}}(\rho=5^{\prime})$& 1.447 &1.54& $\mu$Jy beam$^{-1}$ \\
\hline
\end{tabular}
\end{table}

\section{Catalogue}
\label{sec:cat_cat}

\subsection{Initial compilation}
\label{sec:initial}

\begin{figure}
\centering
\includegraphics[scale=.56,natwidth=6in,natheight=14in]{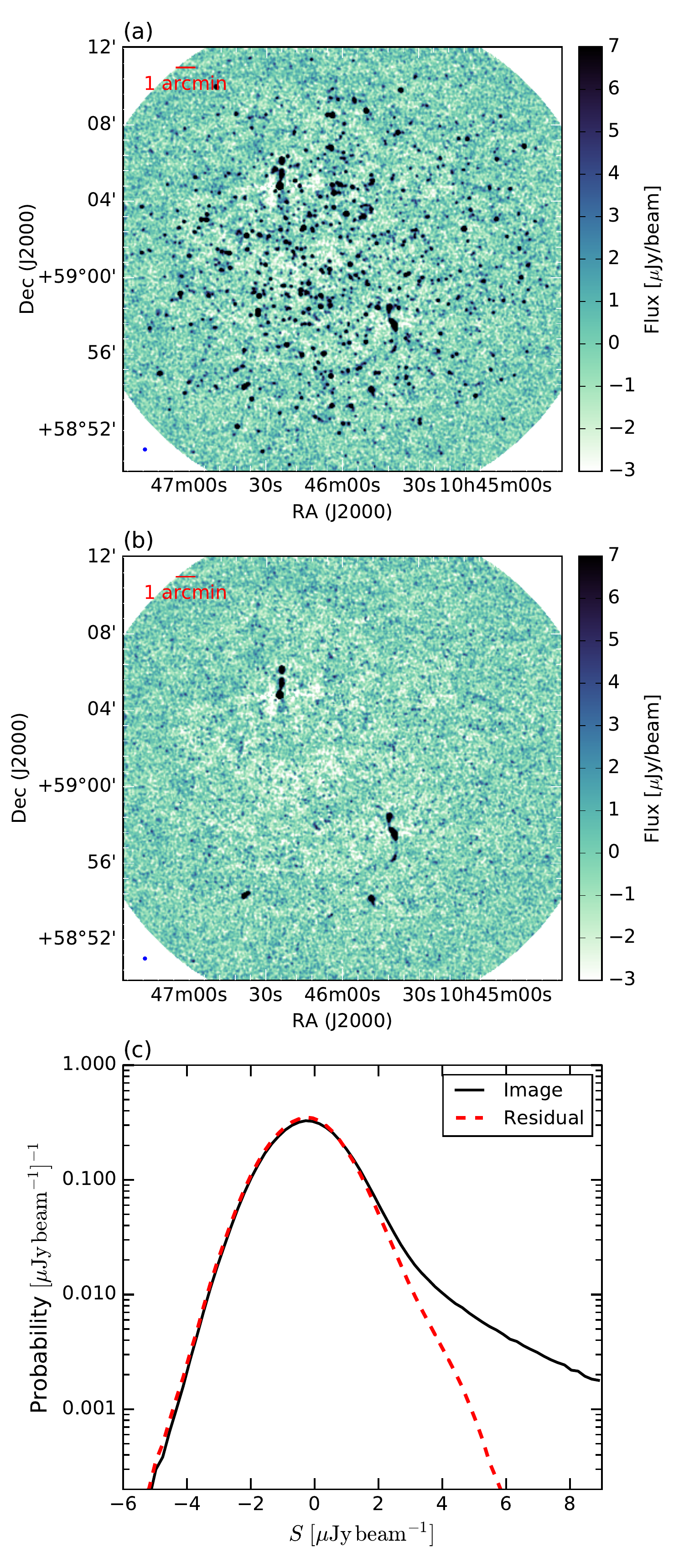}
\caption{VLA 3-GHz Lockman Hole images. The size of the field extends to the $10\,$per cent power radius of the primary beam. Panel (a) shows the field before any primary beam correction and with all sources. Panel (b) shows the residual image after the fitted catalogue sources have been subtracted (again with no primary beam correction), with the exception of the four extended sources not fit by Gaussian models. Panel (c) shows the pixel distributions from each image, with the solid black line being the full image and the red dashed line the residual image. }
\label{fig:ims_res}
\end{figure}

Although images at both resolutions were available, we have constructed a catalogue based on the C-configuration only ($8\,$arcsec resolution) at this time. This is due to the lower instrumental noise of the C-configuration data, as well as the increased sensitivity to extended emission. The higher resolution data were only used here to aid in the cataloguing process.

The initial source lists were obtained in a similar manner as described for the simulations in \citetalias{Vernstrom16a}. The {\Ob} task \textsc{FndSou} was used to locate peaks down to the $3\sigma$ level in each image. In this case the value of $\sigma$ was calculated locally around each peak and no primary beam correction was applied, so it varies across the image with an average value of $\sigma=1.3\, \mu$Jy beam$^{-1}$ (including instrumental and confusion noise). Each peak was then fit with a 2D elliptical Gaussian model, with the fit parameters being the peak flux density \speak, the RA and Dec positions, the major axis FWHM \thetamaj, minor axis FWHM \thetamin, and position angle \posang. The fitting was constrained such that the major and minor axes could not be less than the image synthesized beam size \thetabeam, which is $8\,$arcsec for the C image and $2.75\,$arcsec for the CB image. The total flux density, \stot, was computed as 

\begin{equation}
\label{eq:cat_sint}
S_{\rm total}=S_{\rm peak} \frac{\theta_{\rm maj} \theta_{\rm min}}{\theta_{\rm B}^2}.
\end{equation}
The deconvolved major and minor axes, {\thetadmaj} and \thetadmin, were computed as $\theta_{\rm Dmaj}=\sqrt{\theta_{\rm maj}^2-\theta_{\rm B}^2}$ and $\theta_{\rm Dmin}=\sqrt{\theta_{\rm min}^2-\theta_{\rm B}^2}$.

From the initial component list only sources with fitted peaks above $4\sigma$ were kept. Two residual images were created at the lower resolution: (1) using the results obtained from fitting the low resolution C image; and (2) using the results obtained from fitting the higher resolution CB image but with the parameters adjusted to match those of the lower resolution (i.e. de-convolving the fitted sizes and computing new convolved sizes with the larger beam). We then went through the sources one by one and identified four types of source for follow up.
\begin{enumerate}[label=(\arabic*),leftmargin=*]
\item Complex objects (multi-component sources or ones not well fit by simple Gaussian models). The treatment of these sources is discussed further in Sec.~\ref{sec:ct_srcsz}.
\item Single sources, where the residual image made using the high resolution fit parameters was better (lower peak residual or lower $\chi^2$ value in a box size of two beam widths around the source). In these cases only, the high resolution fit parameters (convolved to the lower resolution) were adopted for the catalogue.
\item Sources fit as a single source in the low resolution image, but multiple sources in the high resolution image. 
\item Sources fit as a single source in the low resolution image with no match, or only one match in the high resolution image, but seemingly poorly fit by a single Gaussian model and with multiple  optical and/or infrared sources within the source area.
\end{enumerate} 
For types 3 and 4, the sources were re-fit manually using a fixed number of Gaussian models (two or three), with initial guesses for the positions taken from the high resolution fits or the positions of the optical/IR counterparts. A new residual image was created and the process was repeated using the residual image, to search for any peaks that were missed by the first round. The \textsc{AIPS} task {RMSD} was used to compute the rms in regions around each pixel. From this image the noise values were found for each source. 

All sources with a signal-to-noise ratio (SNR) less than 5 were removed from the catalogue, leaving the final version with 558 sources. The flux densities were then corrected for the primary beam, based on the beam value at the fitted location. We included sources out to the $12\,$per cent power level of the primary beam; however, any sources past about $50\,$per cent power should be treated with some caution. The full image and final residual image are shown in panels (a) and (b) of Fig.~\ref{fig:ims_res}, with the pixel distributions of the full and residual images shown in panel (c).

For the parameter uncertainties we use equation (41) from \citet{Condon97} for the overall signal-to-noise ratio $\rho$,
\begin{equation}
\scriptsize
\rho^2=\frac{ \theta_{\rm maj} \theta_{\rm min} S_{\rm peak}^2}{4 \theta_{\rm B}^2 \sigma^2}\left [ 1+ \left( \frac{\theta_{\rm B}}{\theta_{\rm maj}} \right )^2 \right ]^{\beta_{\rm maj}}\left [ 1+ \left( \frac{\theta_{\rm B}}{\theta_{\rm min}} \right )^2 \right ]^{\beta_{\rm min}},
\label{eq:cat_jimer3}
\end{equation}
where $\sigma$ is the total rms and $\beta_{\rm maj}$ and $\beta_{\rm min}$ are to be determined by simulations. The units of $\sigma$ and $S_{\rm peak}$ here are Jy beam$^{-1}$. The errors on each of the fit parameters are related to $\rho^2$ by
\begin{equation}
\begin{split}
 \frac{2}{\rho^2} & \simeq \frac{\sigma^2(S_{\rm peak})}{S_{\rm peak}^2}=\frac{\sigma^2(S_{\rm total})}{S_{\rm total}^2} \\ &= 8 \ln{2} \frac{\sigma^2(x_0)}{\theta_{\rm maj}^2} =8 \ln{2} \frac{\sigma^2(y_0)}{\theta_{\rm min}^2}\\& = \frac{\sigma^2(\theta_{\rm maj})}{\theta_{\rm maj}^2}  =\frac{\sigma^2(\theta_{\rm min})}{\theta_{\rm min}^2}  \\  &=\frac{\sigma^2(\phi)}{2} \left ( \frac{\theta_{\rm maj}^2-\theta_{\rm min}^2}{\theta_{\rm maj} \theta_{\rm min}}\right ) .
\end{split}
\label{eq:cat_jimer2}
\end{equation}
Here $x_0$, and $y_0$ are the source position coordinates and $\phi$ is the source position angle. Using the simulations in \citetalias{Vernstrom16a} we determined for $S_{\rm peak}$, $S_{\rm tot}$, $x_0$, and $y_0$ the values $\beta_{\rm maj}=\beta_{\rm min}=1$. However, for the axis sizes, it was discovered in \citetalias{Vernstrom16a} that $\beta$ should vary with axis size and peak flux density, and the adapted equations   
\begin{equation}
\beta_{\rm maj}=k_{\rm maj}-\left ( \frac{\theta_{\rm maj}}{\theta_{\rm B}} \right ) ^2 \log_{10}\left [ \left( \frac{S_{\rm peak}}{\sigma} \right ) ^2 \right],
\label{eq:alphamj}
\end{equation}
and
\begin{equation}
\beta_{\rm min}=k_{\rm min}-\left ( \frac{\theta_{\rm maj}}{\theta_{\rm B}} \right ) ^2 \log_{10}\left [ \left( \frac{S_{\rm peak}}{\sigma} \right ) ^2 \right],
\label{eq:alphamn}
\end{equation}
work well. The values of $k_{\rm maj}$ and $k_{\rm min}$ are determined via simulation to be $k_{\rm maj}=7.47$ and $k_{\rm min}=2.89$. 

\begin{figure}
\includegraphics[scale=.36,natwidth=9in,natheight=9in]{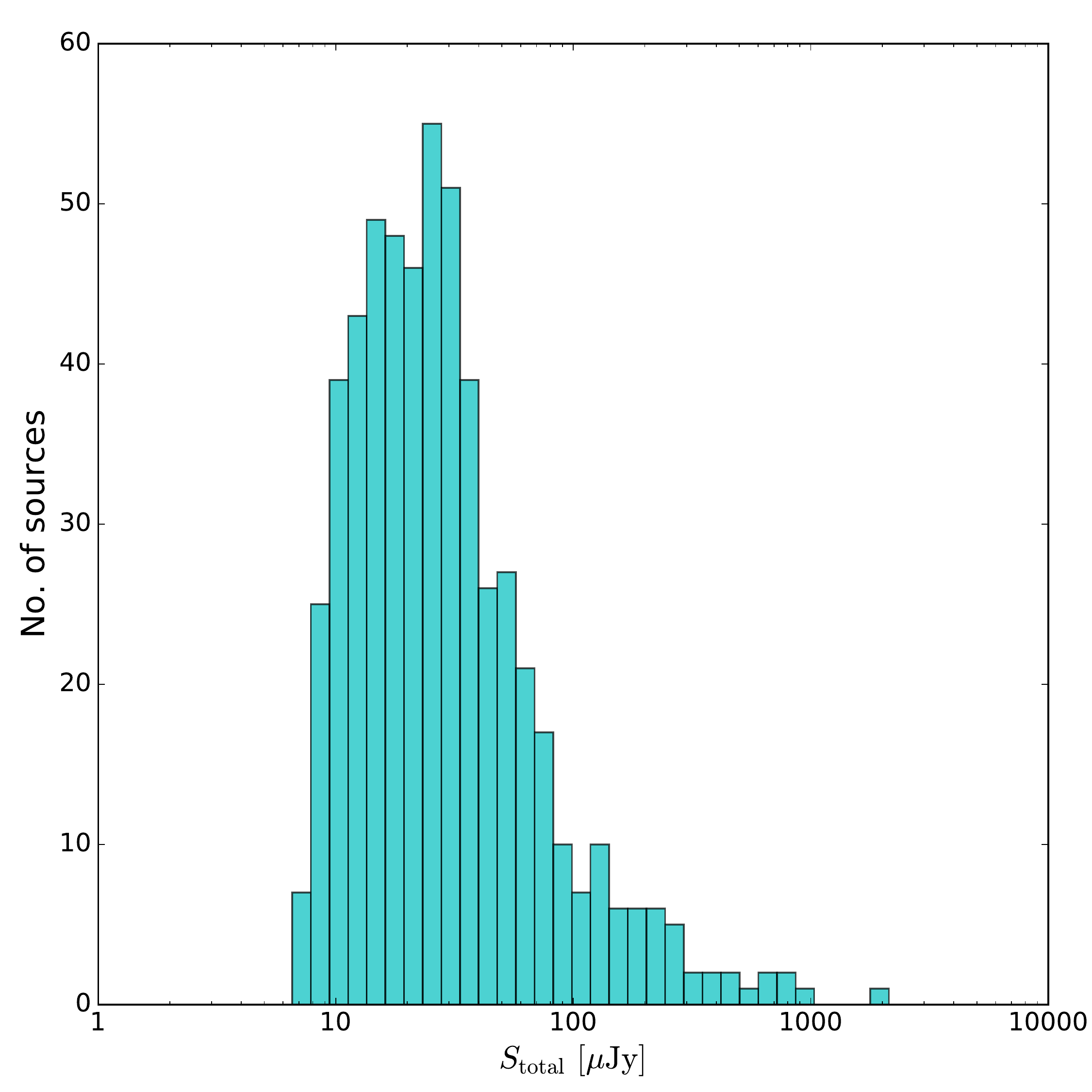}
\caption{Histogram of the total primary beam-corrected flux densities from the catalogue sources on a logarithmic scale.}
\label{fig:cat_flxhist}
\end{figure}

Spectral index estimates for each source are also included in the catalogue, with the methods described further in Sec.~\ref{sec:cat_si}. The flux density distribution is shown in Fig.~\ref{fig:cat_flxhist}. A sample of catalogue entries is shown in  Table~\ref{tab:sampcat}.

\begin{table*}
 \setlength{\tabcolsep}{4.05pt}

\footnotesize
 \caption{
	Lockman Hole North 3-GHz $8\,$arcsec resolution catalogue ordered by Right Ascension. This table has been truncated and is
	available in its entirety in the online version of this paper. The corresponding IAU name is ``TV16J+RA(hhmmss)+Dec(ddmmss)''. For example, the first source is TV16 J104440+585931. We omit the full IDs here to save space, but they are listed in the online catalogue. Columns 1 and 2 give the RA positions and uncertainties, with the uncertainty listed in seconds of time. Columns 3 and 4 give the Dec and associated uncertainties, with uncertainties listed in arcseconds. Column 5 is the effective frequency. Column 6 gives the primary beam attenuation factor. Column 7 lists the uncorrected peak flux densities. Column 8 gives the signal-to-noise ratios. Columns 9 and 10 give the total corrected flux densities and uncertainties. Column 11 is the size flag, where 0 is a non-Gaussian or complex source, 1 is resolved, 2 is partially resolved, 3 is possibly resolved and 4 is unresolved. Columns 12 and 13 are the deconvolved major axis and uncertainties, where a dash indicates an unresolved source. Columns 14 and 15 are the deconvolved minor axises and uncertainties. For columns 12 and 14 an asterisk indicates that the fitted size was less than $2\sigma$ greater than the beam. Column 16 is the deconvolved position angle. Columns 17 and 18 are the measured spectral indices and associated uncertainties (see Sec.~\ref{sec:cat_si}). Here a dash indicates the spectral index is outside the range of $\pm2.25$ or has an uncertainty larger than $0.4$. Only the deconvolved sizes are listed here, however the full version of the catalogue contains both convolved and deconvolved sizes. }
\label{tab:sampcat}
 \begin{tabu}{@{}cccccccccccccccccc@{}}
 \hline
 	%\rowfont{\tiny}
	\multicolumn{1}{c}{RA} &
	\multicolumn{1}{c}{$\sigma_{\trm{\tiny RA}}$} &
	\multicolumn{1}{c}{Dec} &
	\multicolumn{1}{c}{$\sigma_{\trm{\tiny Dec}}$} &
	\multicolumn{1}{c}{$\nu^*$} &
	\multicolumn{1}{c}{$PB$} &
	\multicolumn{1}{c}{$S_{\rm peak}$} &
	\multicolumn{1}{c}{SNR} &
	\multicolumn{1}{c}{$S_{\rm total}$} &
	\multicolumn{1}{c}{$\sigma_{S_{\rm total}}$} &
	\multicolumn{1}{c}{Flag} &
	\multicolumn{1}{c}{\thetadmaj} &
	\multicolumn{1}{c}{$\sigma_{\theta_{\rm Dmaj}}$} &
	\multicolumn{1}{c}{\thetadmin} &
	\multicolumn{1}{c}{$\sigma_{\theta_{\rm Dmin}}$} &
	\multicolumn{1}{c}{\posang} &
	\multicolumn{1}{c}{$\alpha$} &
	\multicolumn{1}{c}{$\sigma_{\alpha}$} \\
	\rowfont{\scriptsize}
	\multicolumn{1}{c}{(J2000.0)} &
	\multicolumn{1}{c}{[sec]} &
	\multicolumn{1}{c}{(J2000.0)} &
	\multicolumn{1}{c}{[arcsec]} &
	\multicolumn{1}{c}{[GHz]} &
	\multicolumn{1}{c}{} &
	\multicolumn{1}{c}{[$\mu$Jy beam$^{-1}$]}&		 
	\multicolumn{1}{c}{} & 
	\multicolumn{1}{c}{[$\mu$Jy]} &
	\multicolumn{1}{c}{[$\mu$Jy]} &
	\multicolumn{1}{c}{size} & 
	\multicolumn{1}{c}{[arcsec]} & 
	\multicolumn{1}{c}{[arcsec]} & 
	\multicolumn{1}{c}{[arcsec]} &
	\multicolumn{1}{c}{[arcsec]} &
	\multicolumn{1}{c}{[deg]} & 
	\multicolumn{1}{c}{} &
	\multicolumn{1}{c}{} \\

	 \rowfont{\scriptsize}
	\multicolumn{1}{c}{(1)} &
	\multicolumn{1}{c}{(2)} &
	\multicolumn{1}{c}{(3)} &
	\multicolumn{1}{c}{(4)} &
	\multicolumn{1}{c}{(5)} &
	\multicolumn{1}{c}{(6)} &
	\multicolumn{1}{c}{(7)} &
	\multicolumn{1}{c}{(8)} &
	\multicolumn{1}{c}{(9)} &
	\multicolumn{1}{c}{(10)} &
	\multicolumn{1}{c}{(11)} &
	\multicolumn{1}{c}{(12)} &
	\multicolumn{1}{c}{(13)} &
	\multicolumn{1}{c}{(14)} &
	\multicolumn{1}{c}{(15)} &
	\multicolumn{1}{c}{(16)}&
	\multicolumn{1}{c}{(17)}&
	\multicolumn{1}{c}{(18)}\\
	\hline
10:44:40.21&0.09&+58:59:31.2&0.7&2.63&5.9&11.4&10.2&69&9.6&3&1.2*&1.7&---&---&---&---&---\\
10:44:42.86&0.20&+58:59:51.1&1.2&2.67&5.6&6.5&6.1&40&9.1&2&6.4&1.3&---&---&$-34$&---&---\\
10:44:44.73&0.08&+59:02:54.3&0.6&2.68&5.6&12.9&11.6&73&8.9&3&0.9*&2.1&---&---&---&---&---\\
10:44:46.87&0.17&+59:01:56.7&1.3&2.73&5&6.5&5.4&32&8.5&4&---&---&---&---&---&---&---\\
10:44:47.54&0.09&+58:59:19.8&0.7&2.73&5.3&11.6&10.2&62&8.6&3&1*&2&---&---&---&---&---\\
10:44:47.58&0.05&+59:00:36.9&0.3&2.74&4.8&24.1&22.4&123&7.7&2&5&0.7&---&---&$-29$&$-1.2$&0.2\\
10:44:47.94&0.12&+59:02:15.9&0.7&2.74&4.3&12.6&9.9&61&8.6&2&6.1&0.9&---&---&$-39$&---&---\\
10:44:48.16&0.08&+58:56:06.5&0.6&2.62&6.7&13&11.7&86&10&3&0.9*&2.1&---&---&---&---&---\\
10:44:48.68&0.21&+59:01:17.2&1.4&2.76&3.8&6.4&5.1&26&7.2&2&4.2&1.1&---&---&$-3$&---&---\\
10:44:48.83&0.01&+59:06:49.4&0.1&2.58&7.7&106&97.5&844&12&4&---&---&---&---&---&$-0.8$&0\\
 \hline
  \hline
 \end{tabu}
\end{table*}

In some cases it is necessary to correct for the effects of time and bandwidth smearing. Bandwidth smearing is an effect of the finite bandwidth and will radially broaden the synthesized beam by convolving it with a rectangle of angular width $\Delta \theta \Delta \nu/\nu$, where $\Delta \theta$ is the radial offset from the pointing centre. With our VLA data, which has $\Delta \nu=2\,$MHz, at the FWHM of the primary beam the bandwidth smearing is roughly $1.0\,$arcsec at $3 \,$GHz. This means that our $8\,$arcsec beam would be convolved radially by a $1\,$arcsec Gaussian yielding a new beam size of $\simeq 8.06\,$arcsec, which is not large enough to cause concern. 

Similarly, time averaging can also affect the synthesized beam. The exact effect of the time smearing depends on the source position and baseline orientation. For a source at the North Pole the effect is simple azimuthal smearing, but for sources away from the pole the effect gets more complicated. Given that our field is reasonably far north, the effect is close to azimuthal. The amount of smearing due to the time resolution goes as $\Delta \theta \Delta t/f$, where $f$ is the Earth's sidereal rotation period of $86164 \, {\rm s}/2\pi=1.37\times 10^4\,$s. Based on \citet{Bridle99}, we estimate the loss in peak flux density of our VLA data, with $\Delta t=1\,$s before any averaging, of only $2\,$per cent for the higher resolution data and less than $1\,$per cent for the lower resolution data at the FWHM of the primary beam. The effect is therefore small enough to be neglected.

\subsection{Sources sizes}
\label{sec:ct_srcsz}

\begin{figure*}
\centering
\includegraphics[scale=0.405,natwidth=9in,natheight=9in]{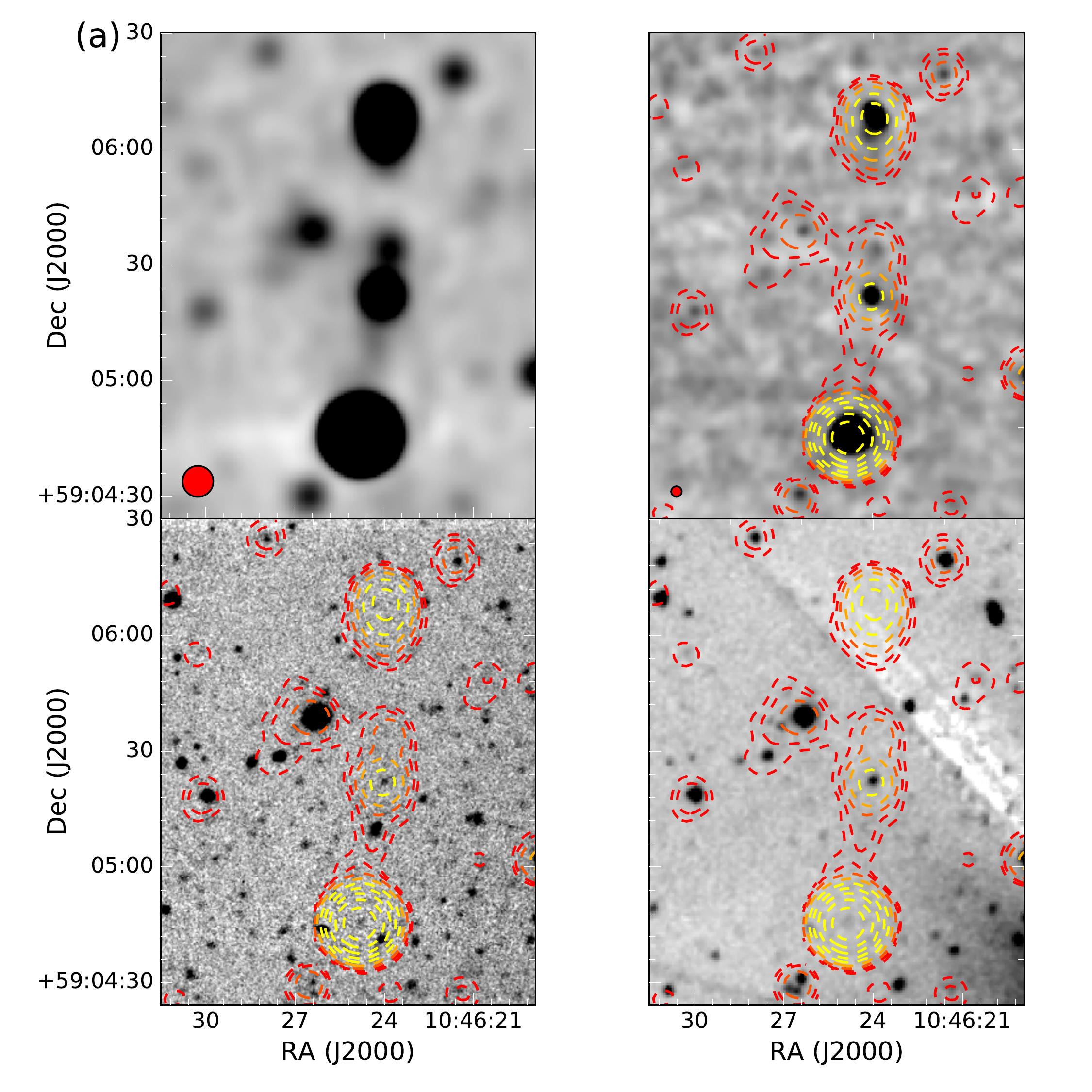}\includegraphics[scale=0.405,natwidth=9in,natheight=9in]{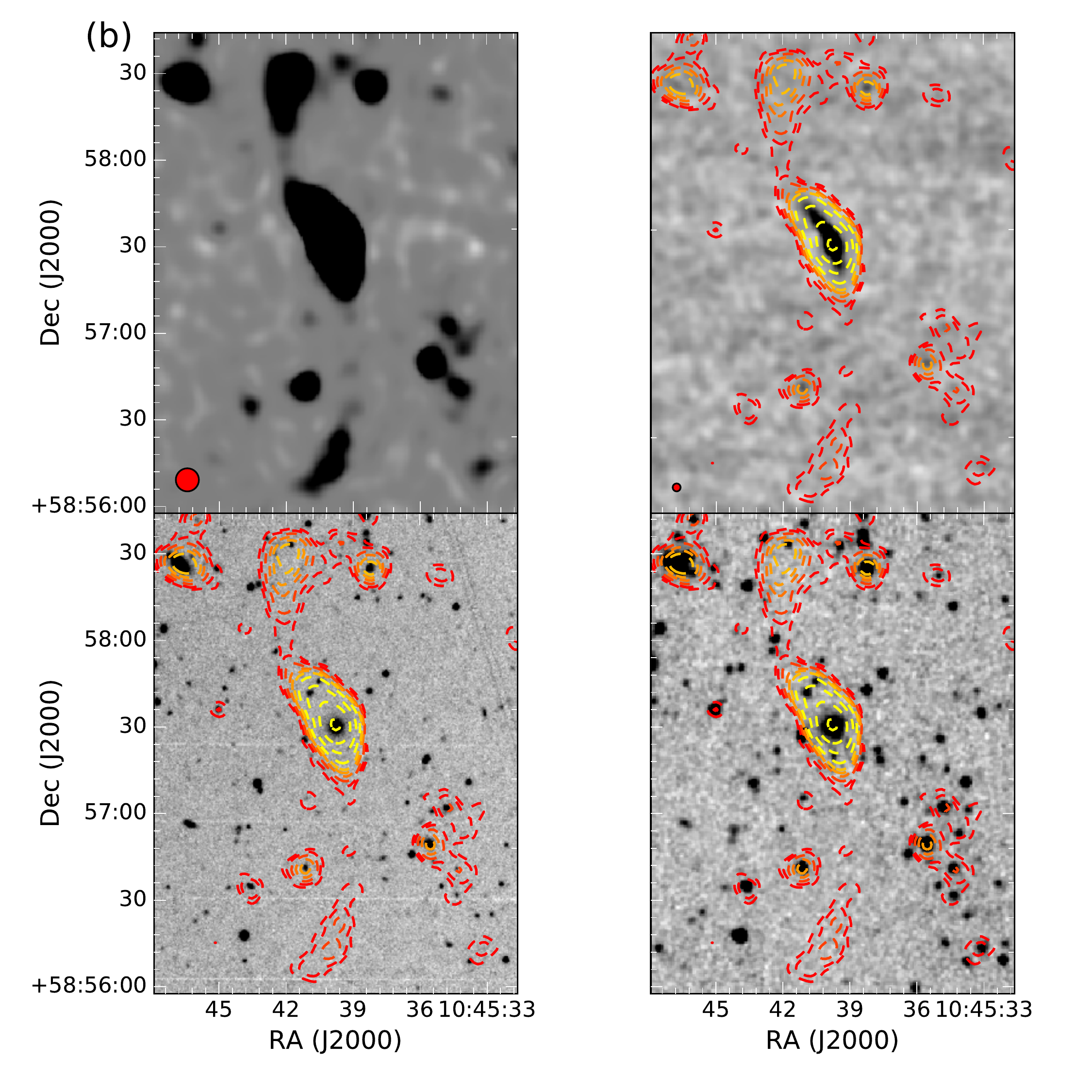}
\includegraphics[scale=0.405,natwidth=9in,natheight=9in]{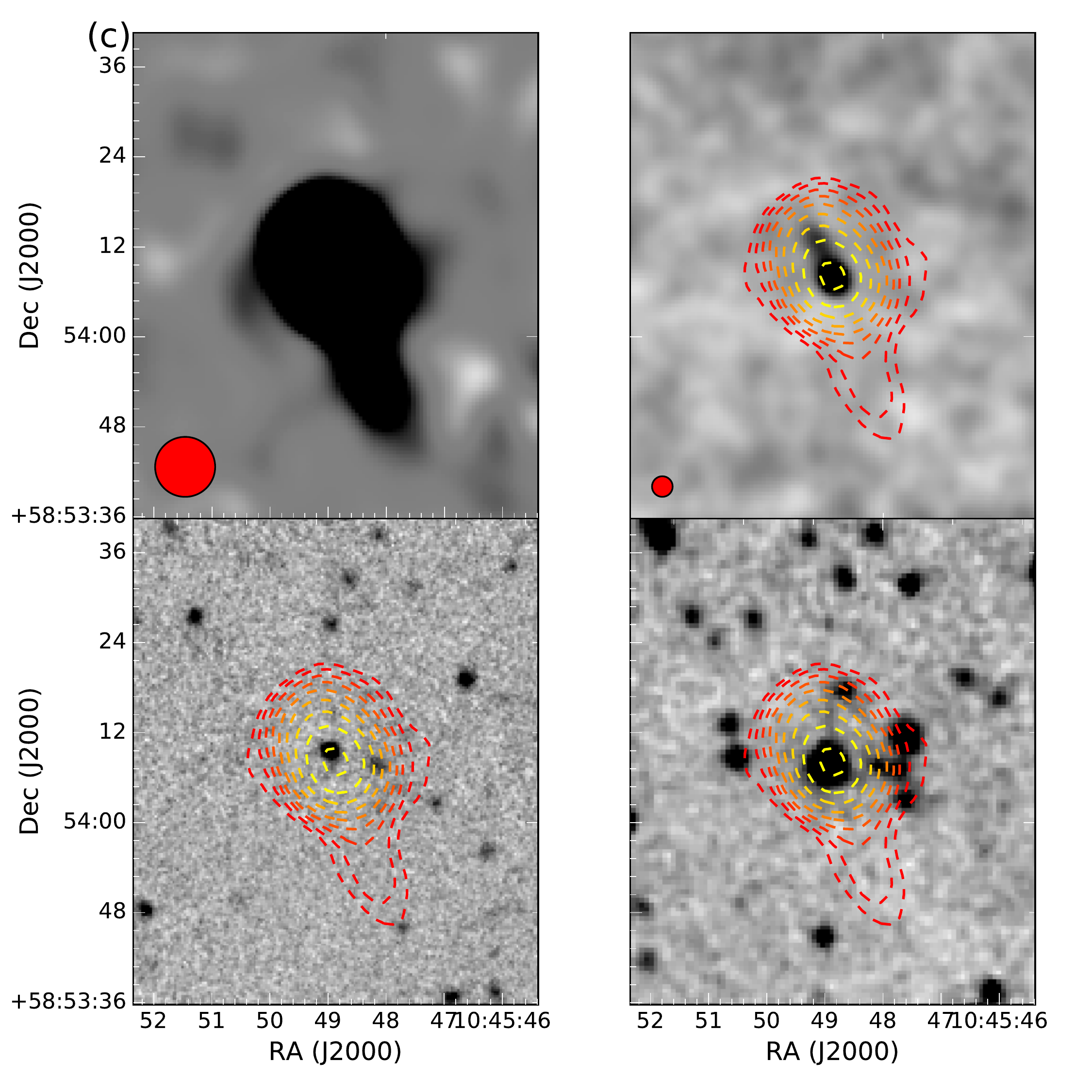}\includegraphics[scale=0.405,natwidth=9in,natheight=9in]{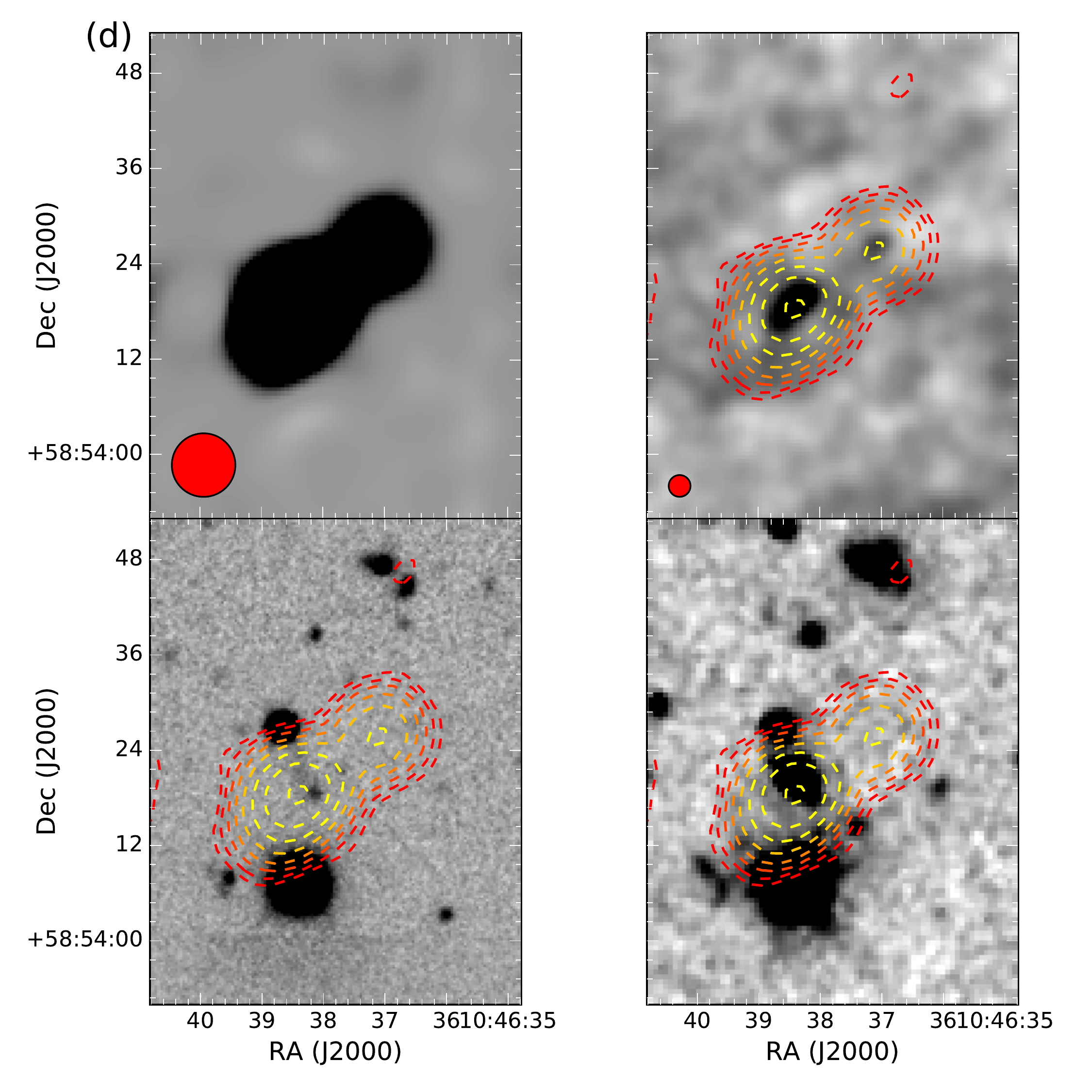}
\caption{Images for the most extended or complex sources in the catalogue. In each group the panels from left to right, top to bottom, are the $8\,$arcsec $3$-GHz image, the $2.75\,$arcsec $3$-GHz image, the KPNO $g$-band image, and the $\textit{Spitzer}$ $3.6$-$\mu$m image. The overlaid contours are from the lower resolution $3$-GHz image. (a): Images for source ID TV16 J104624+590522.   The contours levels are 3, 8, 18, 40, 95, 200, 500, 1100, and 2500 $\sigma$. (b): Images for source ID TV16 J104539+585730. The contours levels are 3, 5, 8, 13, 18, 25, 45, 100, 300, and 720 $\sigma$. (c): Images for source ID TV16 J104548+585408. The contours levels are 3, 6, 9, 15, 25, 35, 60, 90, and 150 $\sigma$. (d): Images for source ID TV16 J104637+585422. The contours levels are 3, 6, 10, 16, 25, 40, 70, and 120 $\sigma$.}
\label{fig:cat_ex4}
\end{figure*}

There are several conventions used in the literature when dealing with source sizes. In this paper, if it is clear that a source has extended structure and is not well fit by a Gaussian, or has multiple components, we consider it to be complex and needs to be treated differently than the situation for compact sources. Axis sizes are only significant if they are at least $2\sigma$ greater than the beam size. If both fitted axes are at least $2\sigma$ larger than the beam these objects are fully resolved. If $\theta_{\rm maj}=\theta_{\rm min}=\theta_{\rm B}$ then these objects are unresolved. If the major axis is resolved, and the minor axis is unresolved, or the minor axis is greater than the beam but less than $2\sigma$ larger than the beam, then the source is partially resolved. If neither axes are resolved, but at least one axis has an upper limit larger than the beam, then the source is considered {\it possibly} resolved. The catalogue includes a size flag to indicate the status. These flags are: (0) complex source; (1) resolved; (2) partially resolved; (3) possibly resolved; and (4) unresolved. The reported sizes include a flag indicating if the size is less than $2\sigma$ greater than the beam. For partially and possibly resolved sources, the total flux is computed as the geometric mean of $S_{\rm tot}$ and $S_{\rm peak}$.

A determination of the source angular size distribution as a function of flux density is needed to interpret radio source counts; however, this distribution is currently not well measured. In particular, it is not known how the size distribution changes with flux density towards faint flux densities, and how it varies by galaxy type. The source size distribution has an effect on the observed source count as well, and it is likely that differences in corrections regarding source sizes are responsible for the large scatter seen in the $1.4$-GHz counts from different surveys in the sub-mJy region \citep{Condon07}. Deep surveys using high-resolution interferometry will tend to miss larger sources near the survey limit, and this may sometimes have been ignored, but at other times over-corrected. 

The C-image resolution is not ideal for investigating this issue, since, with the larger beam, the observed upper limit to source size is $> 1\,$arcsec, particularly for sources near the flux density limit (only 4 sources are fully resolved by our criteria). The average source size (for brighter sources) is estimated to be $0.7\,$arcsec based on high-resolution imaging in \citet{Muxlow07}, although this may be biased towards smaller sources given that the high resolution resolves out larger sources and is likely different for fainter sources. Because of our beam size the catalogue contains fewer sources with small sizes and upper limits, particularly at the faintest flux densities. A more detailed investigation of source sizes will be carried out with the addition of 40 hours of VLA A-configuration data (with a resolution of approximately $0.6\,$arcsec and an instrumental rms $\sigma\simeq 1\, \mu$Jy beam$^{-1}$), which will be discussed in a future paper. 

\subsection{Complex sources }
\label{sec:ct_srccx}

Most of the sources detected by the VLA are compact and are either unresolved or barely resolved. However, there are four sources that are not described by a single Gaussian component, since they are genuinely extended. For each of these four sources the total flux density was found by summing the flux density inside regions guided by contours (but adapted to not include nearby sources). The positions were determined from the peak positions of the radio emission. The reported major and minor axis sizes for these sources are not Gaussian FWHMs, as with the other sources, but rather the square root of the total area used in the flux density summation.

Figure~\ref{fig:cat_ex4} shows the C image for these four sources, as well as the C-image contours superimposed on the CB image, the optical $g$-band image from the Kitt Peak National Observatory \citep[KPNO,][]{Gonzalez11}, and the infrared \textit{Spitzer} $3.6$-$\mu$m IRAC image from the \textit{Spitzer} Wide-area InfraRed Extragalactic survey \citep[SWIRE, ][]{Lonsdale03}, all obtained through the SWIRE image cutout service.\footnote{\url{http://irsa.ipac.caltech.edu/data/SPITZER/SWIRE/index_cutouts.html}} 

The source in panel (a) of Fig.~\ref{fig:cat_ex4} (ID TV16 J104624+590522) is a known quasar, classified as a double-lobed FR \rom{2} source \citep{Veron10}, with a spectroscopic redshift of $z=3.63$. It also has associated X-ray emission seen with {\it Chandra} \citep[X-ray ID CXOX J104623.9+590522,][]{Wilkes09}. The other three sources do not correspond with any known AGN or QSO objects in existing catalogues. 

It is clear from looking at the optical and IR images that there could be emission from multiple objects contributing to the radio images. Regardless, none of these sources are fit well by single or even multiple Gaussian models; therefore we classify them as ``complex''. As stated above, the source in panel (a) of Fig.~\ref{fig:cat_ex4} has a clear optical counterpart near the centre, which is likely associated with the radio emission seen in the centre and the two lobes. The source in panel (b) of Fig.~\ref{fig:cat_ex4} (ID TV16 J104539+585730) is also coincident with a bright optical/IR counterpart near the centre; the extended trailing emission seen in the radio images is likely associated with that object and not other nearby optical/IR objects. 

The source in panel (c) of Fig.~\ref{fig:cat_ex4} (ID TV16 J104548+585408) has an optically bright counterpart near the centre. None of the optical/IR objects line up well with the C-image contours, which show emission to the sides and below the central peak. The extended radio emission is probably from a narrow-angle tail of an AGN. Similarly for the source in panel (d) of Fig.~\ref{fig:cat_ex4} (ID TV16 J104637+585422), the radio emission is likely associated with one main optical object; with the extended emission coming from AGN activity with a prominent knot. Both would be potentially FR \rom{1} type sources, although higher resolution images are needed to investigate the morphology in more detail.

\section{Source Count}
\label{sec:cat_count}

\begin{figure}
\includegraphics[scale=.36,natwidth=9in,natheight=9in]{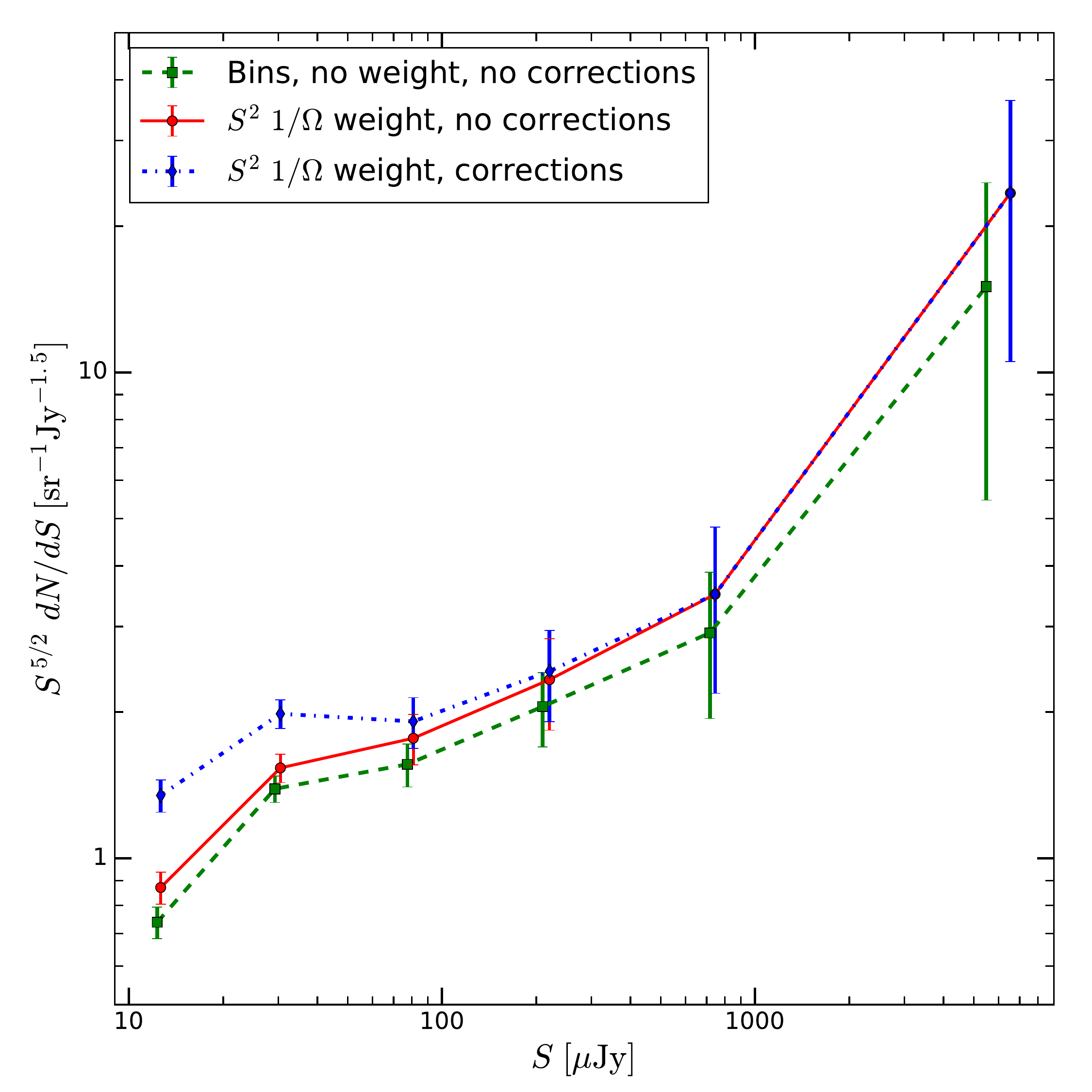}
\caption{Euclidean-normalized $3$-GHz source count corrections. The green dashed line and points is the count computed in bins using equation~(\ref{eq:scc}), with no weights and no corrections for completeness and false detections. The red dot-dashed line and points is the count computed using equation~(\ref{eq:scc3}), with $S^2$ and $1/\Omega$ weighting, but no corrections for completeness and false detections. The blue solid line and points is the count computed using equation~(\ref{eq:scc3}), but with additional corrections for completeness and false detections from the \citetalias{Vernstrom16a} simulations.}
\label{fig:cat_cnt_cors}
\end{figure}

\begin{figure}
\includegraphics[scale=.55,natwidth=6in,natheight=12in]{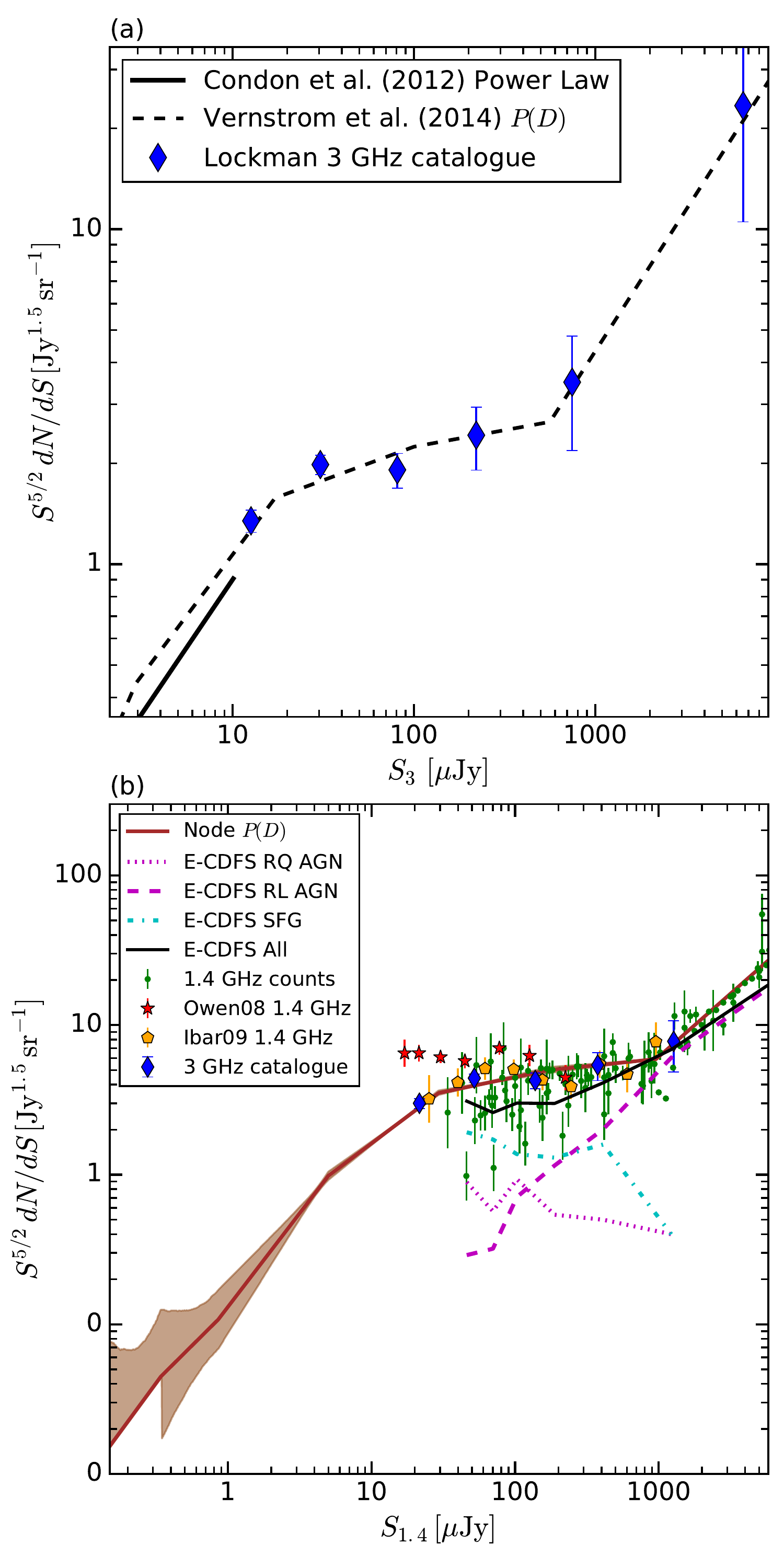}
\caption{Euclidean-normalized catalogue source counts. The blue diamonds are the count using the best-fit total flux densities from our catalogue. Panel (a) shows the $3$-GHz catalogue counts.The black dashed line is the \textit{P(D)} count from \citet{Vernstrom13}, while the black solid line is the power-law \textit{P(D)} from \citet{Condon12}.  The counts from the catalogue are binned (with the points plotted at the mean flux densities), whereas the \textit{P(D)} counts are for a power-law model (or multiple connected power laws). Panel (b) shows the Euclidean-normalized counts scaled to $1.4\,$GHz using a spectral index of $-0.7$. The brown solid line is the {\it P(D)} count with the shaded area being the $68\,$per cent confidence region. The green points are compiled counts from \citet{Dezotti09}. The red stars are the count from \citet{Owen08} in the Lockman Hole North and the orange pentagons are the count from \citet{Ibar09} in other regions of the Lockman Hole. The black solid line is the $1.4$-GHz count from the Extended {\it Chandra} Deep Field South \citep[E-CDFS,][]{Padovani15}. The light blue dash-dotted line represents the star-forming galaxy (SFG) count from E-CDFS. The purple dotted line shows the radio-quiet AGN and the purple dashed line shows the radio-loud AGN from E-CDFS. }
\label{fig:cat_cnt}
\end{figure}

In general the simplest way to construct the differential source count, $dN/dS$, is 
\begin{equation}
{\frac{dN}{dS_a}}=\frac{n_a}{\Omega_a\Delta S_a},
\label{eq:scc}
\end{equation}
where $n_a$ is the total number of sources in bin $a$, $\Delta S_a$ is the width of the bin, and $\Omega_a$ is the area over which the sources with mean flux $\langle S_a \rangle$ could be detected given the primary beam, noise, and detection limit. The angle over which a source can be detected is
\begin{equation}
\Theta_a=\sqrt{\frac{\ln{P_a}}{\ln{2}}} \frac{\theta_{\rm FWHM}}{2},
\label{eq:area_theta}
\end{equation}
where $\theta_{\rm FWHM}$ is the FWHM of the primary beam in radians. $P_a$ is the signal-to-detection limit ratio (rather than just the signal-to-noise ratio). In the simplest case $P_a=\langle S \rangle/(5\sigma)$, where $\langle S \rangle$ is the mean flux density of bin $a$. More accurately $P_a$ will be different for each source, depending on its flux density. If the source is unresolved then $S=S_{\rm peak}=S_{\rm total}$, but if the source is partially or totally resolved then $S=S_{\rm peak}$. In that case the area is calculated as
\begin{equation}
\Omega_a=\pi \Theta_a^2.
\label{eq:area_omega}
\end{equation}
Since this area is different for each source, the inverse of the area solid angle becomes the weight for each source and equation~(\ref{eq:scc}) becomes
\begin{equation}
 {\frac{dN}{dS_a}}=\frac{1}{\Delta S_a}\sum_{S_i=S_a}^{S_i=S_{a+1}} \frac{1}{\Omega(S_i) }.
\label{eq:scc2}
\end{equation}

However, binning causes bias and magnifies statistical errors when the number of objects per infinitesimal flux density range within the bin varies significantly across the bin. In binned counts, the only information about a source that survives is its weight (the inverse solid angle over which it could be detected). Thus $7$-$\mu$Jy sources are counted with $17$-$\mu$Jy sources, which are much less numerous and have much lower weights, and the result is called the count at $12\, \mu$Jy. 

This problem can be minimized by counting a quantity that does not change much across the flux density bin. For example, if the differential source count is closer to $dN/dS\propto S^{-2}$ (which is approximately the case for the faint flux density region), then assigning a weight of $S^2$ for each source and taking a weighted average over the bin to calculate $S^2 dN/dS$ would yield a more meaningful source count. In this case equation~(\ref{eq:scc2}) becomes
\begin{equation}
 {\frac{dN}{dS_a}}=\frac{1}{\Delta S_a \langle S_a^2 \rangle}\sum_{S_i=S_a}^{S_i=S_{a+1}} \frac{S_i^2}{\Omega(S_i) }.
\label{eq:scc3}
\end{equation}
We used equation~(\ref{eq:scc3}) to derive a source count using the best-fit primary-beam-corrected flux densities.

The correction for flux density bias (or flux boosting), as well as completeness and false detection rate estimates, have been performed using simulation data and the 2D interpolation corrections described in \citetalias{Vernstrom16a}. Figure~\ref{fig:cat_cnt_cors} compares counts made with equations~(\ref{eq:scc}) and (\ref{eq:scc3}), with and without the simulation corrections. Figure~\ref{fig:cat_cnt_cors} shows that using equation~(\ref{eq:scc}) results in smaller counts at all flux densities compared to using equation~(\ref{eq:scc3}). The simulation corrections for completeness and false detections have the effect of increasing the count for flux densities less than approximately $200\, \mu$Jy.  A more detailed and comprehensive investigation of the source-size completeness correction will be included in the next catalogue release, which will incorporate data from the VLA A-configuration. 

The final corrected $3$-GHz count is shown in panel (a) of Fig.~\ref{fig:cat_cnt}, along with the \textit{P(D)} fits from \citet{Condon12} and \citet{Vernstrom13}. Table~\ref{tab:scount} lists the Euclidean-normalized count. The close agreement between the direct count and the {\it P(D)} fit (Fig.~\ref{fig:cat_cnt}) is a testament to the power of the {\it P(D)} method, which is not sensitive to the many corrections (such as completeness or size) needed for the catalogue-based count.

Figure~\ref{fig:cat_cnt} (b) shows the catalogue and {\it P(D)} counts scaled to $1.4\,$GHz using a single spectral index of $-0.7$, compared with the previous Lockman Hole counts at $1.4\,$GHz \citep{Owen08}, compiled published counts from \citet{Dezotti09}, and the counts and population source count models from the Extended {\it Chandra} Deep Field South \citep[E-CDFS,][]{Padovani15}. Our current count is currently the only published count at $3\,$GHz and is currently the deepest source count near $1.4\,$GHz. Our (scaled) count falls in line with the spread of previously published counts at $1.4\,$GHz, but deviates from the faint Owen $\&$ Morrison counts (the previously deepest $1.4$-GHz count with an rms of $2.7\, \mu$ Jy beam$^{-1}$, discussed in more detail in Sec.~\ref{sec:cat_owenm}). The count is higher than that from E-CDFS at all flux densities (except near $1000\, \mu$Jy). The count matches well with the count from \citet{Ibar09}, which surveyed other regions of the Lockman Hole at a resolution of $5\,$arcsec with a sensitivity of around $ 6 \, \mu$Jy beam$^{-1}$.

\begin{table}
\centering
\caption{Differential Euclidean-normalized source count.}
\label{tab:scount}
\begin{tabular}{llll}
\hline
\hline
\T \B
$S_{\rm low}$ & $S_{\rm high}$ & $\langle S \rangle$  & $dN/dS\, S^{5/2}$ \\

[$\mu$Jy] &[$\mu$Jy]&[$\mu$Jy]& [Jy$^{1.5}$ sr$^{-1}$] \\
\hline
7 & 17  &13& $\phantom{0}1.3 \pm 0.09$  \T \B \\
17 & 50  &30 &$\phantom{0}1.9 \pm 0.13$ \\
50 & 135  &81& $\phantom{0}1.9 \pm 0.23 $  \\
135 & 400 &220&  $\phantom{0}2.4 \pm 0.52 $ \\
400 & 1500  &750& $\phantom{0}3.5\pm 1.30$  \\
1500 & 8900 &6500 & $23.4 \pm 9.87$  \\
\hline
\end{tabular}
\end{table}

\section{Spectral Indices}
\label{sec:cat_si}

\begin{figure}
\includegraphics[scale=.35,natwidth=9in,natheight=9in]{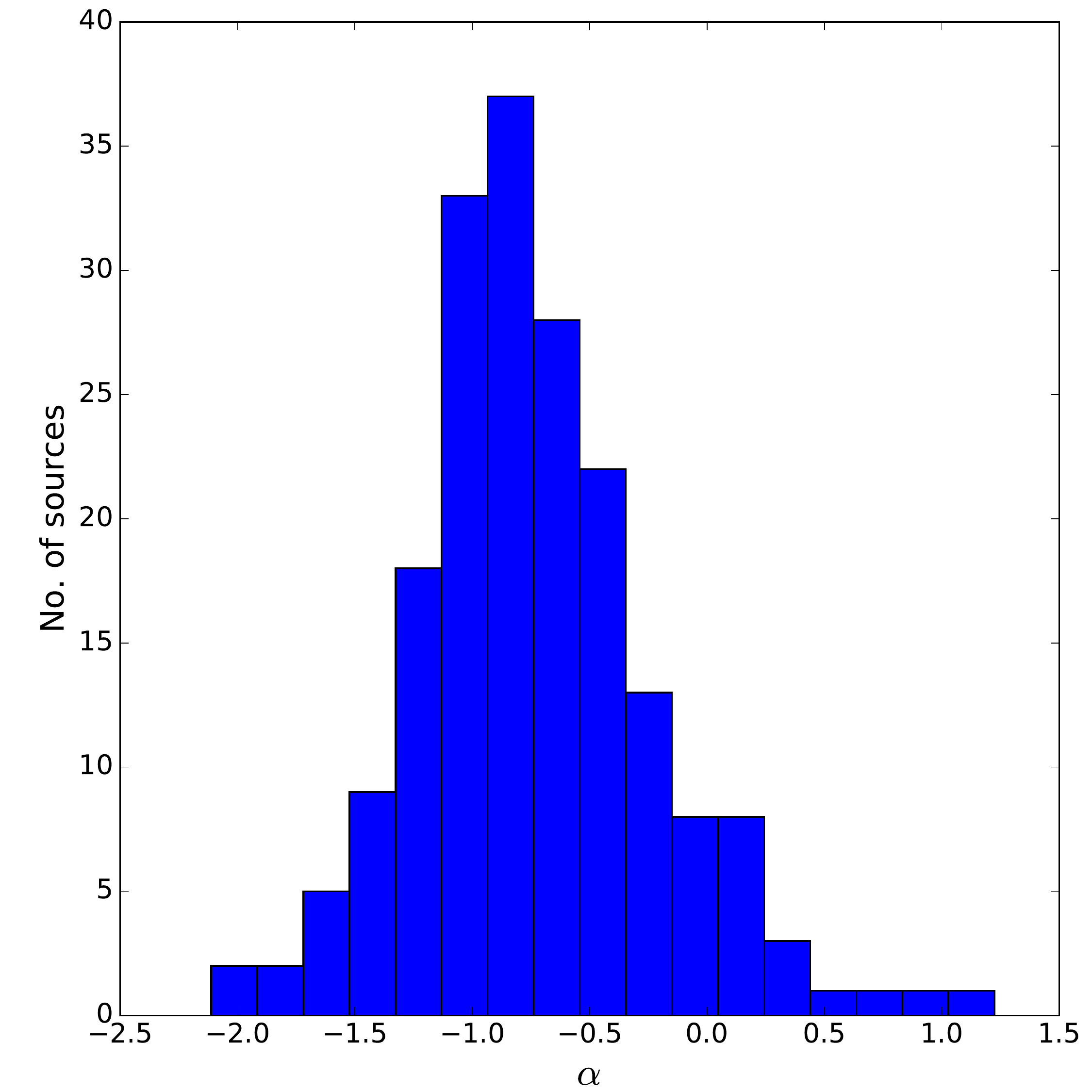}
\caption{Spectral index histogram. The median of the distribution is $\langle\alpha \rangle=-0.76\pm0.04$.}
\label{fig:cat_sihists}
\end{figure}

\begin{figure}
\includegraphics[scale=.47,natwidth=7in,natheight=13in]{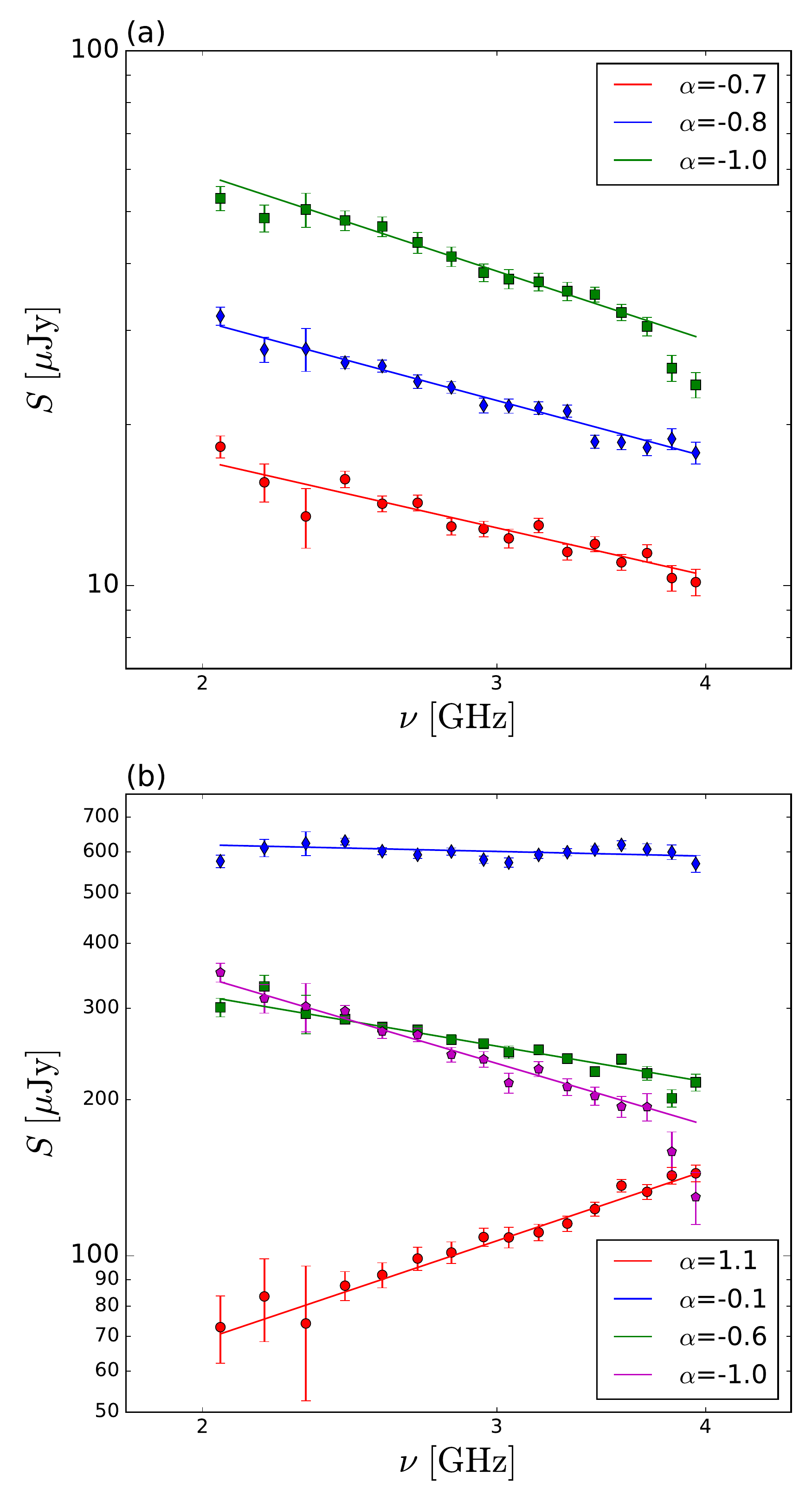}
\caption{Example source spectra. The points are the primary beam-corrected sub-band image values at the fitted peak locations. The solid lines are the best-fit power-law models. Panel (a) shows the points and models for stacking source spectra for sources with $\sigma_{\alpha}>0.3$ or sources where the individual spectral fit was $> |2.25|$ . These were divided into three flux bins: $6\le S \le15$ (red); $15\le S \le 25$ (blue); and $25\le S \le 110$ (green). Panel (b) shows some example individual source spectra: The blue points and line are for a flat-spectrum source with $\alpha=-0.07$ (source ID TV16 J104551+590841); the red points and line are an inverted spectrum source with $\alpha=1.1$ (source ID TV16 J104557+590041); the green points and lines are a negative slope power-law source with $\alpha=-0.6$ (source ID TV16 J104633+585901); and the magenta is an ultra-steep spectrum (USS) source with $\alpha=-1.0$ (source ID TV16 J104537+585730).}
\label{fig:cat_sie}
\end{figure}

%\begin{figure}
%\includegraphics[scale=.37,natwidth=9in,natheight=9in]{fig10_2.pdf}
%\caption{Spectral index image of the double-lobed FR \rom{2} AGN, source ID TV16 J104624+590522. This clearly shows the shallower spectral index of the compact core, with $\alpha_{\rm core} \simeq -0.3$, and the steeper indices in the lobes, $\alpha_{\rm lobe} \simeq -1.2$. The contours are from the lower resolution $3$-GHz image with levels of 3, 8, 18, 40, 95, 200, 500, 1100, and 2500 $\sigma$.  }
%\label{fig:agn_si}
%\end{figure}

The $2$-GHz bandwidth of the VLA allows us to obtain information on the spectral behaviour of the sources, at least for the brighter ones. We first retrieved the flux density values of each source found within the $12\,$per cent beamwidth level at the positions of the fitted peaks in each of the 16 sub-band images. We applied a primary beam correction to each value, based on the primary beam for each sub-band frequency and source position. We then performed a weighted least squares fit for the spectral index $\alpha_i$ and the corresponding fitted $3$-GHz flux density, with the weights derived from the normalized sub-band weight from the sub-band rms and the primary beam correction at the source position.

The uncertainty in the spectral index is $\sigma_{\alpha}\simeq 5/{\rm SNR}$ \citep{Condon15}, where SNR is the peak signal-to-noise ratio. Only sources with spectral indices inside the range $-2.25\le \alpha \le 2.25$ and with $\sigma_{\alpha} \le 0.3$ were included in the catalogue. These criteria leave 140 sources with spectral index estimates. The spectral index distribution is shown in Fig.~\ref{fig:cat_sihists}. The distribution has a median of $\langle\alpha \rangle=-0.76\pm0.04$, in the range expected if the majority of sources are star-forming galaxies. 

For the sources that are excluded by the above criteria, we did a weighted stacked of the spectra and fit for the spectral index, which yielded $\alpha=-0.85$. We also stacked these sources in bins of primary beam-corrected peak flux density (taking the average in each sub-band) and fit for spectral indices. These are shown in panel (a) of Fig.~\ref{fig:cat_sie}. The spectral indices from stacking are $-0.7$ for $6\le S_{\rm peak} \le15$, $-0.8$ for $15\le S_{\rm peak} \le25$, and $-1.0$ for $25\le S_{\rm peak} \le110$. The stacking results show a steepening of the spectral slope with increasing flux density. The brightest flux density bin contains many more sources that are further out in the field, therefore having higher noise when taking into account the primary beams, particularly at the higher frequencies. We did take this into account in the weights when fitting for the spectral index, however, regardless of how the weights were determined (or even neglecting some of the lowest signal-to-noise sources) the spectral index remains equal to or steeper than $-1.0$. 

The majority of source spectra are well described by a power law with a spectral index around $-0.7$. However, there are some sources that deviate from this. There are at least five sources with inverted spectra, or positive spectral indices, which are possible GHz-peak spectrum (GPS) sources \citet{ODea91}. They are likely part of the sub-population of high frequency peakers \citep[HFP,][]{Dallacasa00}, which have peaks at frequencies greater than $1\,$GHz. Inverted-spectrum sources are powerful ($P_{\rm 1.4 GHz} > 10^{25}\,$W Hz$^{-1}$) and compact objects with angular sizes smaller than $1\,$arcsec. These objects are usually interpreted as being AGN in an early stage of evolution. There are also 5 to 10 sources with very flat spectra, $\alpha\simeq 0 \pm 0.2$, which could be flat-spectrum radio quasars (FSRQs) or blazars. These flat or inverted spectra are not characteristic of star-formation processes, and thus can be used as a diagnostic to identify AGN \citep{Huynh07}. There are a similar number of ultra-steep spectrum (USS) sources with $\alpha\le -1.0$. USS sources have been successful tracers of high-redshift radio galaxies \citep[HzRGs, e.g.][]{Tielens79,Chambers96b}. Examples of each of these are shown in panel (b) of Fig.~\ref{fig:cat_sie}.

%The source in panel (a) of Fig.~\ref{fig:cat_ex4}, ID TV16 J104624+590522, is the brightest and most well defined multi-component source in the field. The AGN has two clear lobes and a central compact core. As is common with double-lobed FR \rom{2} sources, the core has a much shallower spectral index than the lobes, with the core having $\alpha_{\rm core}\simeq-0.3$. The bottom lobe has a central spectral index of $-1.1$, steeping to $\simeq -1.5$ near the edges. The top lobe ranges from $-2.0$ to $-1.1$. We created a spectral index image using the $2$-GHz of bandwidth. The section of the spectral index image around this source is shown in Fig.~\ref{fig:agn_si}.

\section{Cross-Identifications}
\label{sec:cat_crxid}

Using a maximum likelihood approach \citep{Sutherland92,Ciliegi03a,Chapin11}, we compared the $3$-GHz source positions with other published catalogues. The likelihood ratio, $L_{i,j}$, compares  the distribution of possible counterpart sources in a matching catalogue with that of background sources. The differential density of ``true'' counterparts for radio sources as a function of their separation $r$ and brightnesses (either flux density $S$ or magnitude depending on the type of counterpart catalogue) and a colour $c$ (if more than one band is available) is $n(S,c,r)$, and is defined as

\begin{equation}
n_{\rm c}(S,c,r)\, \, dS \, dc \, dr =q(S,c) \,f(r) \, \, dS \, dc \, dr.
\label{eq:liken1}
\end{equation}
Here $q(S,c)$ is the distribution of brightnesses and colours (in the matching catalogue) of counterparts, and $f(r)$ is the positional probability distribution as a function of separation for counterparts. The positional probability, assuming a symmetric Gaussian distribution, is 
\begin{equation}
f(r) = \frac{r}{\sigma_r^2}e^{-r^2/2\sigma_r^2},
\label{eq:likefr}
\end{equation} 
normalized so that $\int f(r) dr=1$. The normalization for $q(S,c)$ is such that it integrates to the average expected number of counterparts per radio source.

The relative number of expected true counterparts to background objects is the likelihood ratio:
\begin{equation}
L_{i,j}=\frac{n_{\rm c}}{n_{\rm b}}.
\label{eq:likel1}
\end{equation}
Here $n_{\rm b}$ is the differential number density of background objects, defined as
\begin{equation}
n_{\rm b}(S,c,r)\, \, dS \, dc \, dr = 2 \pi r \rho(S,c) \, \, dS \, dc \, dr,
\label{eq:likenb}
\end{equation}
where $\rho(S,c)$ is the surface density of background objects. Multiplying $\rho(S,c)$ by $ 2 \pi r $ gives the number density of objects at a separation $r$ from a radio source. Using this definition for $n_{\rm b}$ turns equation~(\ref{eq:likel1}) into
\begin{equation}
L_{i,j}=\frac{q(S,c)e^{-r^2/2\sigma_r^2}}{2 \pi \sigma_r^2 \rho(S,c)}.
\label{eq:likel2}
\end{equation}
Larger $L_{i,j}$ values indicate a higher probability of a true counterpart match to a radio source.

For more details on the cross-matching procedure used see appendix B of \citet{Chapin11}. This likelihood ratio cross-matching procedure was performed for each of the additional wavelength catalogues described below.  

\subsection{X-ray, Optical, IR, and redshifts}
\label{sec:cat_opt}

\begin{figure}
\includegraphics[scale=.53,natwidth=6.in,natheight=12in]{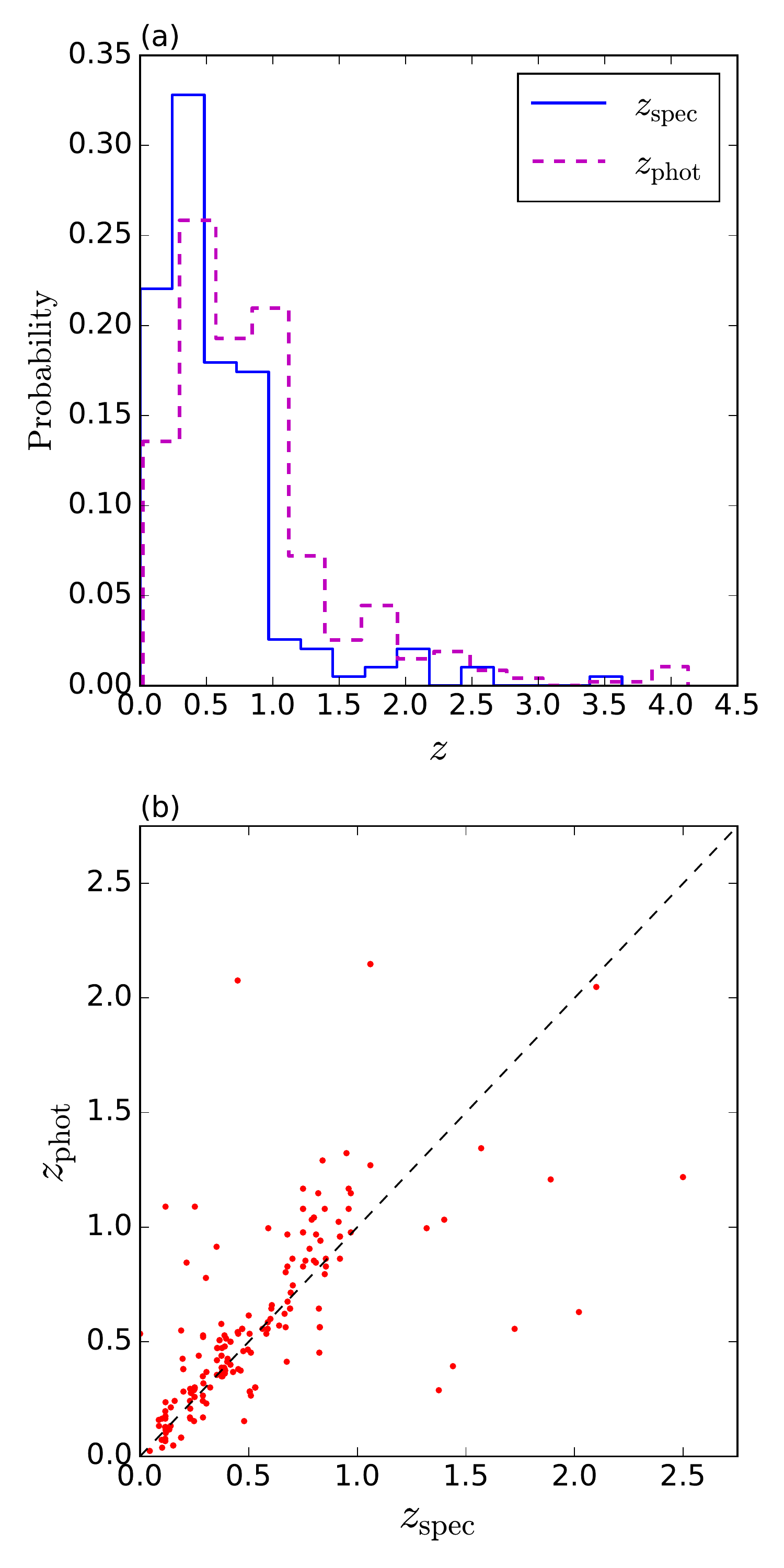}
\caption{Catalogue redshift distributions. Panel (a) is the redshift probability distributions, where the blue solid line shows the distribution of spectroscopic redshifts per unit redshift interval and the purple dashed line is for the photometric redshifts. Panel (b) shows the spectroscopic redshift vs. photometric redshift for those sources with both measurements. The one-to-one value is shown by the black dashed line.}
\label{fig:cat_reddist}
\end{figure}

\begin{figure*}
\includegraphics[scale=.53,natwidth=12in,natheight=12in]{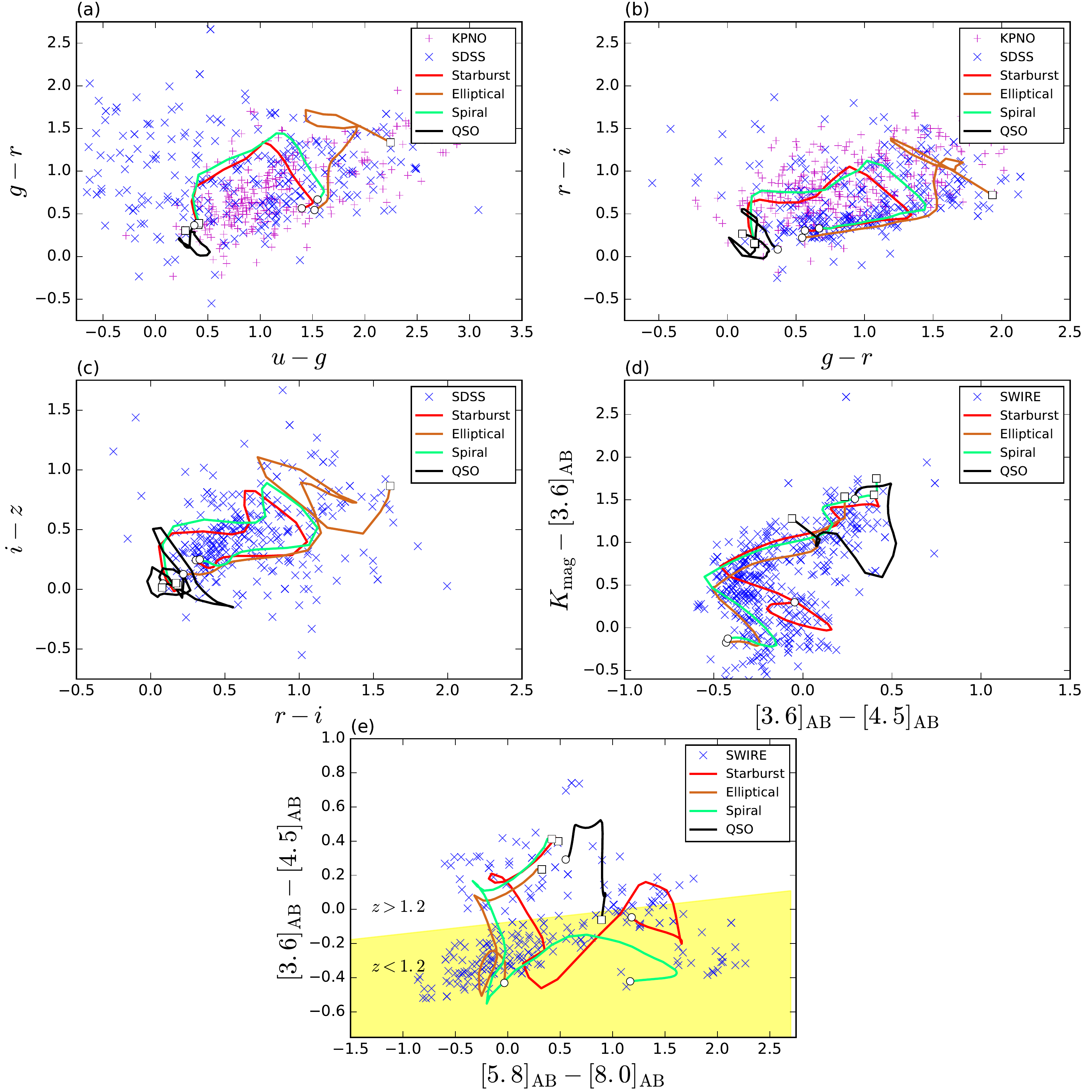}
\caption{Colour-Colour plots of optical and NIR counterpart matches for our radio catalogue. Panels (a) and (b) include data from the SDSS DR9 and INTWFC/KPNO, while panel (c) only includes SDSS. Panel (d) is for \textit{Spitzer} IRAC and UKIDSS data, while panels (e) and (f) are for \textit{Spitzer} IRAC only. The coloured lines are redshift tracks for $0<z<4$ (squares for $z=0$ and circles for $z=4$, assuming no evolution) using galaxy templates from \citet{Polletta07}. The yellow shaded region in panel (e) shows $z<1.2$ from equation~\ref{eq:cat_zline}, as suggested by \citet{Marsden09}.  }
\label{fig:cat_colorc}
\end{figure*}

\begin{figure}
\includegraphics[scale=.47,natwidth=7in,natheight=13in]{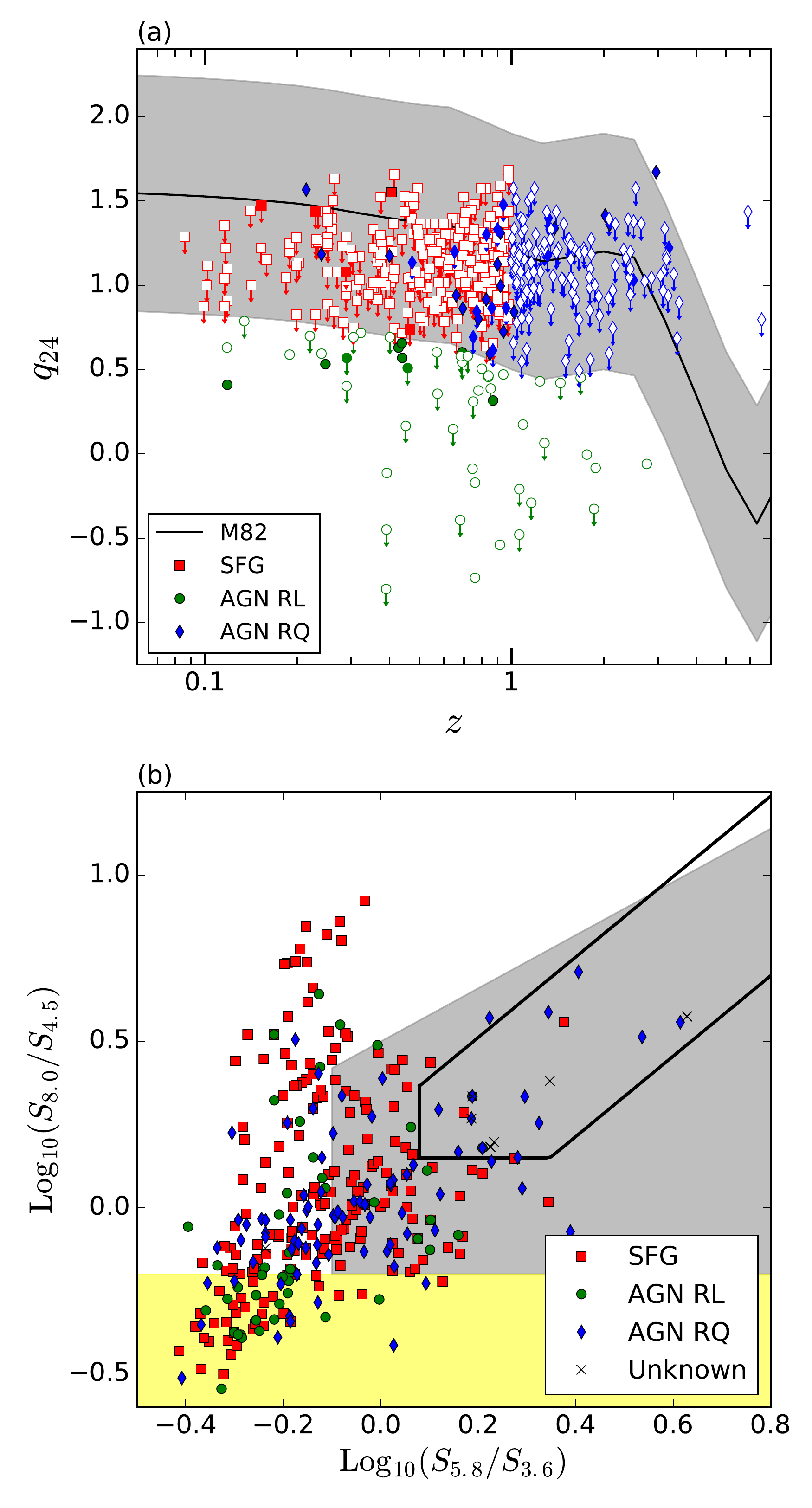}
\caption{Classification criteria plots. For both panels the red squares are sources classified as star-forming galaxies (SFG), the green circles are radio-loud AGN (AGN RL), the blue diamonds are radio-quiet AGN (AGN RQ), and the black crosses are sources without an identification. Panel (a) shows $q_{24}$ as a function of redshift. The black line shows the evolution for M82, with the $\pm 2 \sigma$ region shaded grey. For all the points filled symbols are sources with an X-ray detection, while outlined sources only have an X-ray upper limit; arrows indicate the sources for which there was no $24$-$\mu$m detection and the upper limit was used for $q_{24}$. Panel (b) shows the IRAC colour-colour plot, with the additional AGN classification criteria from \citep[][black dot-dashed lines]{Donley12} and \citep[][yellow region for hot-mode AGN and grey region for cold-mode AGN]{Lacy04,Lacy07,Lacy06a}.}
\label{fig:classify}
\end{figure}

Our VLA source catalogue was cross-matched with several optical and near infrared catalogues. The X-ray data came from the SWIRE {\it Chandra} catalogue \citep{Wilkes09}, which has a limiting broadband (0.3--$8\,$keV) flux of $4\times 10^{-16}\,$erg cm$^{-2}$ s$^{-1}$. For optical data we used the Sloan Digital Sky Survey (SDSS) data release 9 \citep{Ahn12}, which includes $u,g,r,i,$ and $z$ bands with  limits of 22, 22.2, 22.2, 21.3, and 20.5 magnitudes, respectively. Additionally, \citet{Gonzalez11} observed the Lockman Hole with the Isaac Newton Telescope Wide Field Camera (INT WFC) in the $u,g,r,i,$ and $z$ bands, with AB magnitude limits of 23.9, 24.5, 24.0, 23.3, and 22.0, respectively. This survey also included supplementary observations covering the Lockman Hole North in the $g$, $r$, and $i$ bands with the Mosaic 1 camera on the Mayall 4-m Telescope of the Kitt Peak National Observatory (KPNO). 

We additionally used the catalogue from the United Kingdom Infrared Telescope (UKIRT) Infrared Deep Sky Survey (UKIDSS) data release 9 \citep{Lawrence12,Lawrence07}, containing $J$ and $K$ bands, with magnitude depths of 23.1 and 22.5, respectively. Further into the infrared, there are \textit{Spitzer} IRAC and MIPS data from the SWIRE survey at 3.6, 4.5, 5.8, 8.0, and $24 \, \mu$m, with depths of around $10 \, \mu$Jy at $3.6$ and $4.5\, \mu$m (for more details see the SWIRE data release 2 document).\footnote{\url{http://irsa.ipac.caltech.edu/data/SPITZER/SWIRE/docs/delivery_doc_r2_v2.pdf}} Additionally, there are ``warm'' \textit{Spitzer} data from the \textit{Spitzer} Extragalactic Representative Volume Survey (SERVS) at 3.6 and 4.5 $\mu$m, with  a detection limit of approximately $2\, \mu$Jy \citep{Mauduit12}. Furthermore, there are submillimetre data from the {\it Herschel} Multi-tiered Extragalactic Survey third data release \citep[HerMES,][]{Roseboom10,Roseboom12,Oliver12} at 250, 350, and $500\, \mu$m.\footnote{The DR3 HerMES catalogues are composed of SPIRE flux densities extracted at {\it Spitzer} $24$-$\mu$m prior positions. The catalogues were downloaded from \url{http://hedam.lam.fr/HerMES/}.}  Many of these data are available as a precompiled SERVS-matched ``Data Fusion'' catalogue \citep{Vaccari16}.

In addition to multi-wavelength data, we also cross-matched our catalogue with the SWIRE photometric redshift catalogue \citep{Rowan08}, which includes photometric and spectroscopic redshifts. As a secondary redshift catalogue \citep[used when no match was found in][]{Rowan08}, we additionally crossed with the deep SWIRE field catalogue of \citet{Strazzullo10}. The redshift distributions are shown in panel (a) of Fig.~\ref{fig:cat_reddist}. Our catalogue has 184 spectroscopic matches with $\langle z_{\rm spec} \rangle=0.6$ and 456 photometric redshifts with $\langle z_{\rm phot} \rangle=0.9$. These values are similar to those found by \citet{Luch15}, who examined cross-matched radio and optical sources in the Lockman Hole East and found a peak in the distribution near $z=0.7$, with a median redshift of $z=1.04$. Panel (b) of Fig.~\ref{fig:cat_reddist} shows the spectroscopic vs. photometric redshifts for the 181 sources with both measurements. The ratio of spectroscopic to photometric redshifts for those with both values has a mean of $1.03\pm0.06$.

The numbers of matches for each catalogue at each matching band or wavelength are shown in Table~\ref{tab:matches_op} and Table~\ref{tab:matches_ir}.

\begin{table*}
\centering
\caption{Number of source matches in optical catalogues by observing band. The total number of sources in the radio catalogues is 558. This table contains cross-match information from the SWIRE {\it Chandra} survey \citep{Wilkes09}, the Sloan Digital Sky Survey \citep[SDSS,][]{Ahn12}, the Kitt Peak National Observatory \citep[KPNO,][]{Gonzalez11}, and the United Kingdom Infrared Telescope (UKIRT) Infrared Deep Sky Survey (UKIDSS) data release 9 \citep{Lawrence12,Lawrence07}.}
\label{tab:matches_op}
\begin{tabular}{cccccccccccc}
\hline
\hline
 {\it Chandra}&\multicolumn{5}{c}{SDSS}&\multicolumn{4}{c}{KPNO}&\multicolumn{2}{c}{UKIDSS} \\
 &$u$& $g$ & $r$ & $i$ & $z$ &$u$& $g$ & $r$ & $i$ & $J$ &$ K$\\
\hline
51 &302 & 302& 302& 302& 302&287&389&426&394&510&510\\
\hline
\end{tabular}
\end{table*}

\begin{table*}
\centering
\caption{Number of source matches in IR, submm, and redshift catalogues by wavelength. This table contains cross-match information from the \textit{Spitzer} Wide-area InfraRed Extragalactic survey \citep[SWIRE, ][]{Lonsdale03} using both the IRAC and MIPS cameras, the \textit{Spitzer} Extragalactic Representative Volume Survey \citep[SERVS,][]{Mauduit12}, the {\it Herschel} Multi-tiered Extragalactic Survey third data release \citep[HerMES,][]{Oliver12}, and redshifts from the SWIRE photometric redshift catalogue \citep{Rowan08} and \citet{Strazzullo10}. Here $z_{\rm spec}$ is spectroscopic redshift and $z_{\rm phot}$ is photometric. The total number of sources in the radio catalogue is 558. } 
\label{tab:matches_ir}
\begin{tabular}{cccccccccccc}
\hline
\hline
\multicolumn{4}{c}{IRAC} &\multicolumn{2}{c}{SERVS}&MIPS & \multicolumn{3}{c}{SPIRE}&\multicolumn{2}{c}{Redshifts}  \\
$3.6$&$4.5$&$5.8$&$8.0$&$3.6$&$4.5$&$24$&$250$&$350$&$500$&$z_{\rm spec}$ &$z_{\rm phot}$ \\

 [$\mu$m] & [$\mu$m]&[$\mu$m]&[$\mu$m]&[$\mu$m]&[$\mu$m]&[$\mu$m]&[$\mu$m]&[$\mu$m]&[$\mu$m]&&\\
\hline
508&502&345&277&523&502&291&317&283&162&184&456\\
\hline
\end{tabular}
\end{table*}

\subsection{Classification}
\label{sec:classes}

We computed colour cuts for our catalogue using the cross-matched data, which can be seen in Fig.~\ref{fig:cat_colorc}. The SDSS optical data have more scatter than the INT/KPNO data simply because the INT/KPNO data are deeper than SDSS. Additionally, on each plot we have shown four galaxy redshift tracks (starting at $z=0$ going to $z=4$, assuming no evolution) using rest-frame galaxy spectral energy distribution (SED) templates from \citet{Polletta07}. We chose one galaxy from each category of elliptical, spiral, starburst or ULIRG, and quasar. In panel (e) of the Fig.~\ref{fig:cat_colorc}, showing IRAC $[5.8]-[8.0]$ vs $[3.6]-[4.5]$ colours, the yellow region shows the divide between $z<1.2$ and $z>1.2$ from \citet{Marsden09}. The equation for the dividing line is
\begin{equation}
\label{eq:cat_zline}
\left ( [3.6]-[4.5] \right )=0.0682 \times \left ( [5.8]-[8.0] \right )-0.075.
\end{equation}

Using the multi-wavelength and redshift data the sources can be classified into different populations: star-forming galaxies (SFG); radio-loud AGN (RL); and radio-quiet AGN (RQ). To do this we follow the method laid out in \citet{Bonzini13}, who performed analysis with the E-CDFS catalogue. This entails using the $24$-$\mu$m flux density $S_{24}$, the redshift, the X-ray luminosity $L_{\rm X}$, and the IRAC colours. The $24$-$\mu$m flux density is used to compute the mid-IR to radio flux density ratio $q_{24}$, defined as
\begin{equation}
q_{24}=\log_{10}\left [\frac{S_{24}}{S_{\rm r}} \right ].
\label{eq:q24}
\end{equation}
In our case $S_{\rm r}$ is the $3$-GHz flux density. Where no $24$-$\mu$m flux density was available the $24$-$\mu$m detection limit was used to calculate an upper limit. The {\it Chandra} SWIRE hard band flux (2--8 kev) was used along with the redshifts to compute the X-ray luminosity; where there was no detection in the X-ray the detection limit of $4\times 10^{-16}\,$erg cm$^{-2}$ s$^{-1}$ was used as an upper limit.

To classify the sources, we looked at the $q_{24}$ vs. $z$ values and compared those against a star-forming galaxy template. We used the SED of M82 as a template and calculated the $q_{24}$ value at different redshifts by shifting the SED. The average $q_{24}$ spread for star-forming galaxies is $\pm 0.35\,$dex \citep{Sargent10}. The $q_{24}$ values are shown in panel (a) of Fig.~\ref{fig:classify}, along with the M82 template and the $\pm 2 \sigma$ region for the template. 

We also use IRAC colour-selection criteria for AGN from \citet{Donley12} and \citet{Lacy04,Lacy07,Lacy06a}. The IRAC colour-colour plot is shown in panel (b) of Fig.~\ref{fig:classify}, along with the Donley and Lacy selections. For sources without an IRAC detection in all four IRAC bands, the band detection limit was used for the non-detected band(s) to compute the upper limit. The Donley criteria are defined as follows (where $\wedge$ is the logical ``AND'' operator):
\begin{equation}
\begin{split}
x&=\log_{10}\left (\frac{f_{5.8}}{f_{3.6}} \right );\\
y&=\log_{10}\left (\frac{f_{8.0}}{f_{5.8}}\right );
\label{eq:cat_zline1}
\end{split}
\end{equation}
\begin{equation}
\begin{split}
[y\ge (1.21\times x)-0.27]  \, \,\wedge \, \, \, \, \, \, \, \, \, \, \, \, \, \, \, \, \, \, \, \, \,   \, \, \, \, \, \, \, \, \, \, \, \, \, \, \, \, \, \, \, \, \, \, \, \, \, \, \, \, \, \, \, \, \, \, \, \, \, \, \, \,  \, \, \, \, \, \, \,  & \\
[y \le (1.21\times x)+0.27]  \, \, \wedge \, \, \, \, \, \, \, \, \, \, \, \, \, \, \, \, \, \, \,    \, \, \, \, \, \, \, \, \, \, \, \, \, \, \, \, \, \, \, \, \, \, \, \, \, \, \, \, \, \, \, \, \, \, \, \, \, \, \, \, \, \,  \, \, \, \, \, \, \, &\\
\,[x \ge 0.08]  \wedge [y\ge 0.15] \, \, \wedge \, \, \, \, \, \, \, \, \, \, \, \, \, \, \, \, \, \, \,   \, \, \, \, \, \, \, \, \, \, \, \, \, \, \, \, \, \, \, \, \, \, \, \, \, \, \, \, \, \, \, \,\, \, \, \,  \, \, \, \,  \, \, \, \, \, \, \,  &\\
[f_{4.5} > f_{3.6}] \,  \wedge \, [f_{5.8} > f_{4.5}] \,  \wedge \, [f_{8.0}> f_{5.8}] \, \, \, \,\, .
\label{eq:cat_zline2}
\end{split}
\end{equation}
The Lacy selection cuts are $\log_{10}[S_{8.0}/S_{4.5}]> -0.2$, $\log_{10}[S_{5.8}/S_{3.6}]> -0.1$, and $\log_{10}[S_{8.0}/S_{4.5}] \le 0.8 \log_{10}[S_{5.8}/S_{3.6}]+0.5$. The Lacy criteria create three regions that, according to \citet{Luch15}, separate low accretion rate hot-mode AGN (lower yellow section), rapidly accreting cold-mode AGN (upper right grey wedge), and star-forming galaxies (upper and middle left, no colour).

The classification scheme is as follows.
\begin{enumerate}[label=(\arabic*),leftmargin=*]
\item Radio-loud AGN: sources with $q_{24}$ values below the $2\sigma$ lower limit defined by the M82 template. If no redshift was available a mean $q_{24}$ cutoff of $0.7$ (for sources out to a moderate redshift) was adopted as the cutoff, and sources must have $q_{24}$ below this, as well as falling into one of the IRAC AGN regions.
\item Radio-quiet AGN: sources must have a $q_{24}$ value above the $2\sigma$ lower limit, as well as having $L_{\rm X}$ (or an upper limit of $L_{\rm X}$) $\ge 10^{42}\,$erg s$^{-1}$ \citep[The lower limit for AGN-driven sources determined by ][]{Szokoly04}. If no redshift is available, sources must have $q_{24}\ge 0.7$ and fall into one of the IRAC colour AGN regions. 
\item Star-forming galaxies: sources must have a $q_{24}$ value above the $2\sigma$ lower limit, as well as having $L_{\rm X}$ (or an upper limit of $L_{\rm X}$) $< 10^{42}\,$erg s$^{-1}$. If no redshift is available, sources must have $q_{24}\ge 0.7$ and fall outside of the IRAC colour AGN regions.
\item Unknown: sources with too few IRAC band detections and no redshift are unclassifiable. Also, when sources have redshifts and $q_{24}$ values above the cutoff, with $L_{\rm X}<$ cutoff, but are inside the IRAC AGN regions (or vice versa) there is no information to distinguish between SFG and RQ AGN.
\end{enumerate}

Using these criteria we found 58 radio-loud AGN ($10\,$per cent), 155 radio-quiet AGN ($28\,$per cent), 325 SFGs ($58\,$per cent), and 20 unknown ($4\,$per cent). The radio-quiet AGN contribution should definitely be considered as an upper limit, given that the main criterion for separating them from SFGs is the X-ray luminosity, for which a large number had non-detections and the $L_{\rm X}$ value used for classification is based on the detection limit. The optical and IR colours, combined with the template tracks and other classification criteria, show that the majority of sources are likely to be spiral or star-forming types of galaxies at intermediate redshift. This is not unexpected. From the source count and luminosity functions, it is believed that star-forming galaxies dominate in the sub-mJy regime. However, the sub-mJy regime is an area where the contribution from the different populations is known to change dramatically and the exact contribution from each population is still an area of active investigation. \citet{Bonzini13} found $19\,$per cent radio loud AGN, $24\,$per cent radio quiet AGN, and $57\,$per cent SFG with E-CDFS data ($1.4\,$GHz frequency, resolution of $\sim2.2\,$arcsec, and a sensitivity of approximately $7\, \mu$Jy beam$^{-1}$). \citet{Ibar09} found $25\,$per cent radio-quiet AGN in other regions of the Lockman Hole ($1.4\,$GHz frequency, resolution of $5\,$arcsec, and sensitivity of approximately $6 \, \mu$Jy beam$^{-1}$). Both of these estimates are in close agreement with our findings.

\subsection{Radio}
\label{sec:cat_owenm}

\begin{figure}
\includegraphics[scale=.5,natwidth=6.5in,natheight=12in]{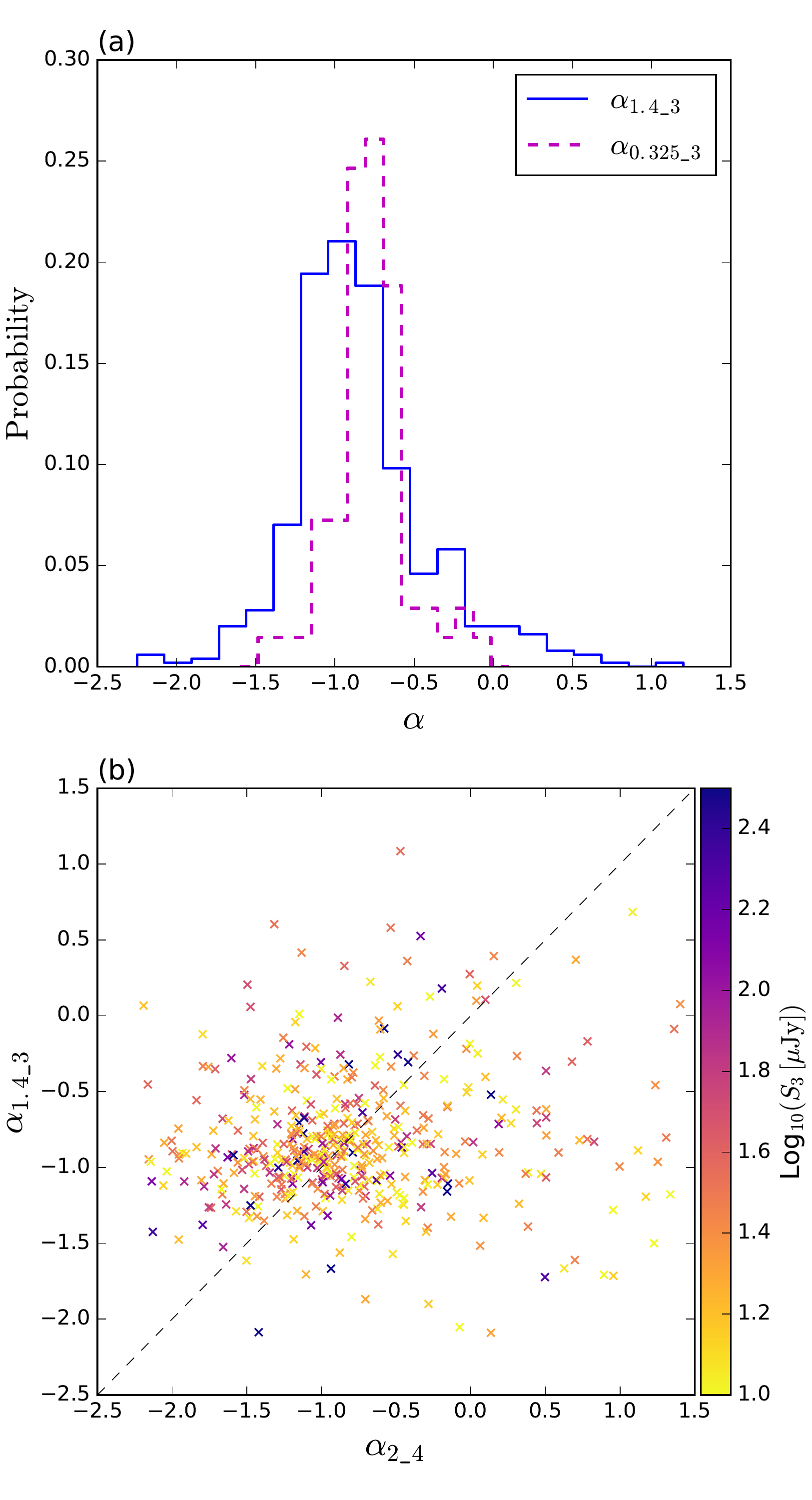}
\caption{Results of cross-matching with the \citet{Owen08} $1.4$-GHz catalogue and \citet{Owen09} $325$-MHz catalogue. Panel (a) shows the $1.4\,$GHz to $3\,$GHz spectral index probability distribution (solid blue line) and the $325\,$MHz to $3\,$GHz spectral index distribution (dashed purple line). Panel (b) shows $\alpha_{2-4}$ vs $\alpha_{1.4-3}$ with the colour scaling as the logarithm of the $3$-GHz flux densities. The one-to-one value is shown by the black dashed line. }
\label{fig:cat_owensim}
\end{figure}

We cross-matched our VLA catalogue with the $1.4$-GHz VLA catalogue from \citet{Owen08}, finding 499 matches. The data from \citet{Owen08} has a depth of $2.7\, \mu$Jy beam$^{-1}$, with a beam size of $1.6\,$arcsec. The majority of the $1.4$-GHz sources that did not have matches in our catalogue were near the Owen $\&$ Morrison signal-to-noise limit or were near the edge of our cataloguing area, where the sensitivity of the $3$-GHz image is worse.

We computed the spectral indices of the matched sources, finding the distribution shown in panel (a) of Fig.~\ref{fig:cat_owensim}. The average spectral index is approximately $\langle \alpha \rangle=-0.85\pm0.02$, which is steeper than the expected $-0.7$ or indeed the values discussed in Sec.~\ref{sec:cat_si}. Panel (b) of Fig.~\ref{fig:cat_owensim} also shows our internal spectral indices compared against those from cross-matching with the $1.4$-GHz data.

The cause of this discrepancy is not clear. Panel (b) of Fig.~\ref{fig:cat_owensim} does not show any correlation of higher $\alpha_{1.4-3}$  with flux density (the colour scaling of the points), ruling out the possibility that the differences are simply due to faint sources. The average deconvolved size for the $1.4$-GHz sources is smaller than the average deconvolved size at $3\,$GHz; thus to obtain a steep spectral index with our source sizes being larger would mean that either we underestimated our flux densities on average or that \OW over-estimated the $1.4$-GHz flux densities. This is particularly curious because the $1.4$-GHz image had a resolution of approximately $1.6\,$arcsec and simulations, such as those done in \citetalias{Vernstrom16a}, have shown that fits for sources with sizes extending to the beam size or larger tend to underestimate the total flux density, whereas it would seem these source flux densities are {\it over}estimated. However, we believe it most likely that the $1.4$-GHz flux densities are overestimated.

We additionally cross-matched our catalogue with the $325$-MHz VLA catalogue \citep{Owen09}, which has an rms $70\, \mu$Jy beam$^{-1}$ with a $6\,$arcsec beam size. Only 69 sources had matches with the existing catalogue at this frequency for the region. The spectral index distribution for $\alpha_{0.325-3}$ is also shown in panel (a) of Fig.~\ref{fig:cat_owensim}. The mean spectral index is $\langle \alpha_{0.325-3} \rangle=-0.76\pm0.03$, which is similar to the mean $\alpha$ from our 2- to $4$-GHz bandwidth.

\section{Conclusions}
\label{sec:cat_conc}
We have presented the details of the $3$-GHz source catalogue from a single VLA pointing in the Lockman Hole North. The catalogue was made at a resolution of $8\,$arcsec, with instrumental noise rms of approximately $1.01\, \mu$Jy beam$^{-1}$ and a $5\sigma$ detection threshold. The catalogue contains 558 sources. 

We have used the wide bandwidth of the upgraded VLA to obtain spectral index estimates based on independent measurements typically made at widely spaced frequencies; these agree with previous estimates for the mean spectral index of $\mu$Jy sources of $\langle \alpha \rangle \simeq -0.76\pm0.04$. Results from stacking spectra in bins of flux density show evidence for a flattening of the spectral with decreasing flux density.   

We calculated the source count, which was found to be in good agreement with the {\it P(D)} estimate from \citet{Vernstrom13}, providing evidence for the power of the {\it P(D)} technique as an independent means of estimating source counts (moreover, the {\it P(D)} technique can constrain the count to much deeper levels). The count is within the scatter of most previous source counts from other surveys at $1.4\,$GHz after scaling with $\alpha=-0.7$; the most obvious exception is the low flux density count of \citet{Owen08}). Using corrections for source size and completeness, as determined from simulations in \citetalias{Vernstrom16a}, our Euclidean-normalised count shows no signs of the levelling off below the $50$-$\mu$Jy regime seen by \citet{Owen08}. We were able to match a majority of our catalogue sources with those from the \citet{Owen08} $1.4$-GHz catalogue, and the steepness of the spectral indices suggests that the earlier $1.4$-GHz flux densities were systematically overestimated.

Our catalogue was cross-matched with X-ray, optical, infrared, and redshift catalogues, finding a mean redshift around $z=1$. Optical and IR colours were obtained and compared with model galaxy templates as well as multiple AGN population selection criteria. Using a detailed classification scheme yields an upper limit of $10\,$per cent as radio-loud AGN, an upper limit of $28\,$per cent radio-quiet AGN, and $58\,$per cent as star-forming galaxies, with $4\,$per cent unclassifiable; this is in agreement with previous estimates. A separate table with all of the cross-match values is available as supplementary material along with the electronic version of this paper.

Further study of this field and the $\mu$Jy source population will be carried out using additional VLA data in the A configuration. This will yield a detailed comparison of the same sources at multiple resolutions, with comparable noise, allowing for a more in-depth analysis of the source size distribution as a function of source brightness and population type. This should help to resolve the issue of source count discrepancies due to size corrections in the sub-mJy regime. Additionally, a detailed investigation of the far-infrared-to-radio correlation would help to further categorize the sources, as well as to provide information on the evolution of the FIR-to-radio correlation (and the star formation history) as a function of redshift.  

\section{Acknowledgments}
The Dunlap Institute is funded through an endowment established by the David Dunlap family and the University of Toronto. We acknowledge the support of the Natural Sciences and Engineering Research Council (NSERC) of Canada.  We thank the staff of the VLA, which is operated by the National Radio Astronomy Observatory (NRAO), a facility of the National Science Foundation operated by the Associated Universities, Inc. This research has made use of data from the HerMES project (\url{http://hermes.sussex.ac.uk/}). HerMES is a {\it Herschel} Key Programme utilizing Guaranteed Time from the SPIRE instrument team, ESAC scientists and a mission scientist. The HerMES data were accessed through the Herschel Database in Marseille (HeDaM - \url{http://hedam.lam.fr}) operated by CeSAM and hosted by the Laboratoire d'Astrophysique de Marseille. HerMES DR3 was made possible through support of the Herschel Extragalactic Legacy Project, HELP. HELP is a European Commission Research Executive Agency funded project under the SP1-Cooperation, Collaborative project, Small or medium-scale focused research project, FP7-SPACE-2013-1 scheme.

\bibliographystyle{mnras}

\label{lastpage}
\end{document}